\documentclass[%
pre,preprint,superscriptaddress,tightenlines,
showpacs,
showkeys,
a4paper,12pt
]{revtex4}
\usepackage{amssymb,amsmath}
\usepackage{graphicx}

 \renewcommand{\Re}{\mathop{\rm Re}\nolimits}
 \newcommand{\Tr}{\mathop{\rm Tr}\nolimits}
 \newcommand{\T}{\textit{T}}
 \newcommand{\bs}[1]{\boldsymbol{#1}}
 \newcommand{\vc}[1]{\mathbf{#1}}
 \newcommand{\uvc}[1]{\mathbf{\hat #1}}
 \newcommand{\ubs}[1]{\boldsymbol{\hat #1}}
 \newcommand{\dd}{\mathrm{d}}
 
 \newcommand{\eff}{\mathrm{eff}}

\begin{document}
 \DeclareGraphicsExtensions{.jpg,.pdf}

 \title{\bfseries Light scattering by 
optically anisotropic scatterers I: \\ 
 \T--matrix theory for radial and uniform anisotropies}

\author{A.D.~Kiselev}
\email[E-mail: ]{kisel@elit.chernigov.ua}
\affiliation{%
 Chernigov State Technological University,
 Shevchenko Street 95,
 14027 Chernigov, Ukraine
} 
\author{V.Yu.~Reshetnyak}
\email[E-mail: ]{reshet@iop.kiev.ua}
\affiliation{%
 Kiev University,
 Prospect Glushkova 6,     
 03680 Kiev, Ukraine
} 
\author{T.J.~Sluckin}
\email[E-mail: ]{t.j.sluckin@maths.soton.ac.uk} 
\affiliation{%
 Faculty of Mathematical Studies, 
 University of Southampton,
 Southampton, SO17 1BJ, UK
 }

\date{\today}

\begin{abstract}
 
We extend the \T-matrix approach to light scattering by spherical
particles to some simple cases in which the scatterers are optically
anisotropic.  Specifically we consider cases in which the spherical
particles include radially and uniformly anisotropic layers. We find
that in both cases the \T-matrix theory can be formulated using a
modified \T-matrix ansatz with suitably defined modes.  In a uniformly
anisotropic medium we derive these modes by relating the wave packet
representation and expansions of electromagnetic field over spherical
harmonics.  The resulting wave functions are deformed spherical
harmonics that represent solutions of the Maxwell equations.  We use
these modes to express the equations for the \T-matrix elements in
terms of computationally tractable coefficient functions.

\end{abstract}



\maketitle


\section{Introduction}
\label{sec:intro}

The problem of light scattering by particles of one medium embedded in
another has a long history, dating back almost a century to the
classical exact solution due to Mie~\cite{Mie:1908}. The Mie solution
applies to scattering by uniform spherical particles with isotropic
dielectric properties. More recently this strategy has been
successfully applied to ellipsoidal particles and some circumstances
in which the dielectric tensor is anisotropic
~\cite{Asan:1975,Rot:1973,Arag:1990,Arag:1994,Kar:1997,Kis:2000:biano,
Kis:2000:opt}.

Unfortunately, Mie--type solutions, although exact, are not always
physically meaningful. An alternative strategy useful in the long
wavelength limit is the so-called Rayleigh-Gans (R-G) approximation
\cite{New}, which is closely allied to the classic Born approximation
of quantum mechanics. However, detailed information on the precise
range of validity of this approximation often requires solution of the
Mie problem, which is only available in some specific cases.

There are a large number of physical contexts in which it is useful
to understand light scattering by  impurities~\cite{Ishim}. A
particular example of recent interest concerns liquid crystal
devices. There are now a number of systems in which liquid crystal
droplets are suspended in a polymer matrix -- the so-called PDLC
systems -- or the inverse system, involving  colloids now with a
nematic liquid crystal solvent. These inverse  systems are commonly 
known as filled nematics~\cite{Kre:1996,Bel:1998}.

In such systems one needs to calculate light  scattering  by
composite anisotropic particles embedded in an isotropic or an
anisotropic matrix. The model scatterer usually consists of a small
central isotropic particle (``the core''), coated by a much larger
region in which the optical tensor is anisotropic. This is equivalent
to examining light scattering by a composite particle consisting of
the central core plus a surrounding liquid crystalline layer.

The analysis of a Mie--type theory uses a systematic expansion of
the electromagnetic field over vector spherical harmonics.  The
specific form of the expansions is known as the \T--matrix ansatz.
This has been widely used in the related problem of light scattering
by nonspherical particles~\cite{Mis:1996,Mish}.

Recently in~\cite{Kis:2000:biano,Kis:2000:opt} we have studied the
scattering problem for the optical axis distributions of the form:
$n_r\uvc{r}+n_{\vartheta}\ubs{\vartheta}+n_{\varphi}\ubs{\varphi}$.
By using separation of variables and expansions over vector spherical
harmonics, we have developed the generalised Mie theory as an
extension of the \T--matrix ansatz~\cite{Mis:1996}.  This theory combines
computational efficiency of the \T--matrix approach and well defined
transformation properties of the spherical harmonics under rotations.

In this paper we discuss this theory in more detail and explain how
this approach can be extended to the case of uniformly anisotropic
spherical particles.

The layout of the paper is as follows.  General discussion of the
model is given in Sec.~\ref{sec:model}.  Then in
Sec.~\ref{sec:t-matrix-approach} we outline the \T-matrix formalism
for the isotropic medium in the form suitable for subsequent
generalisation.  In Sec.~\ref{sec:scatt-from-radi}, as the simplest
case to start from, we consider how the \T--matrix ansatz applies for
the radially anisotropic layer.  We find that the structure of
electromagnetic modes in the layer requires modification of the
standard \T-matrix ansatz.  In addition, we detail computing the
elements of \T-matrix.
 
In Sec.~\ref{sec:scatt-uni-anis} we describe the
method to put the scattering problem into the language of \T--matrix
by linking the representations of plane wave packets and of spherical
harmonics.  For uniformly anisotropic scatterer we define 
generalised spherical harmonics and show that the effect of
angular momentum mixing can be treated efficiently.

Finally in Sec.~\ref{sec:concl} we draw together the results and make
some concluding remarks.  In particular, we emphasise the importance
of anisotropy effects by making comparison between angular
distributions of scattered wave intensities for radially anisotropic
layer and effective isotropic layer of the same scattering efficiency.
  
Details  on some  technical  results  are relegated to 
Appendices~\ref{sec:vect-spher-harm}~--~\ref{sec:coef-func}. 

\section{Model}
\label{sec:model}

We consider scattering by a spherical particle of radius $R_1$
embedded in a uniform isotropic dielectric medium with dielectric
constant $\epsilon_{ij} = \epsilon \delta_{ij}$ and magnetic
permeability $\mu_{ij}=\mu \delta_{ij}$.  The scattering particle
consists of an inner isotropic core of radius $R_2$, surrounded by an
anisotropic annular layer of thickness $d=R_1-R_2$.       
 
Within the inner core of the scatterer
the dielectric tensor $\bs{\epsilon}$, and the magnetic
permittivity  $\bs{\mu}$ take the values
$\bs{\epsilon}_{ij}=\epsilon_2 \delta_{ij}$, 
$\bs{\mu}_{ij}=\mu_2\delta_{ij}$.
The dielectric tensor within the annular layer is locally uniaxial.
The optical axis distribution is defined by the vector field
$\uvc{n}$. (Hats will denote unit vectors.)
Then within the annular layer
$\bs{\epsilon}_{ij}({\bf r})=\epsilon_1 \delta_{ij}+\Delta\epsilon_1
(\uvc{n}({\bf r})\otimes\uvc{n}({\bf r}))_{ij} $ and 
$\bs{\mu}_{ij}=\mu_1\delta_{ij}$. The unit vector $\uvc{n}$
corresponds to a liquid crystal director for material within the
annular region.

\begin{figure*}[!thb]
\centering
\resizebox{120mm}{!}{\includegraphics{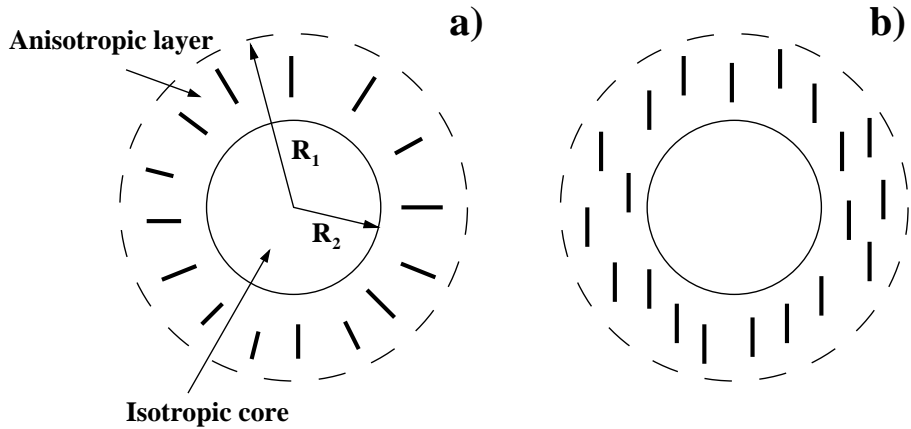}}
\caption{
Distributions of optical axis in the anisotropic layer
around a spherical particle
for radial and uniform structures: 
a)~$\uvc{n}=\uvc{r}$; b)~$\uvc{n}=\uvc{z}$
}
\label{fig:str}
\end{figure*}

We shall suppose that the director field can be written in one of the
following forms
\begin{subequations}
  \label{eq:conf}
\begin{gather}
\uvc{n}=\uvc{z}
\label{eq:unif}\\
  \uvc{n}=\uvc{r}
\label{eq:rad}
\end{gather}
\end{subequations}
where 
$\uvc{r}=(\sin\theta\cos\phi, \sin\theta\sin\phi, \cos\theta)$
is the unit radial vector;
$\phi$ and $\theta$ are Euler angles of the unit vector $\uvc{r}$.
$\uvc{x}$, $\uvc{y}$ and $\uvc{z}$ are the unit vectors
directed along the corresponding coordinate axes.

In Fig.~\ref{fig:str} we have shown these director
distributions. 
Fig.~\ref{fig:str}a shows the radial (and spherically symmetric) director
distribution.
The scattering problem in this case
has already been discussed by Roth and Digman~\cite{Rot:1973}, and in this
case the spherical symmetry of the problem plays an important role in
rendering the Maxwell equations soluble. 

We show in Fig.~\ref{fig:str}b  the case in which 
the optical axis  is directed along the $z$--axis and is
uniformly distributed within the annular layer; this is given by Eq.~\eqref{eq:unif}. 
The case where the scatterer is a long cylinder parallel to $\uvc{n}$
presents no difficulties and can be treated in
cylindrical coordinate system~\cite{Bor}.
Scattering from spherical uniformly anisotropic particles 
is not exactly soluble~\cite{Bor} and has been
studied by using the Rayleigh--Gans method and the anomalous diffraction
approximation in~\cite{Zum:1986,Zum:1988}.  

A simple limit of the physical situations we consider puts $\epsilon
=\epsilon_1 =\epsilon_2$, with $\mu =\mu_1 =\mu_2$. The first
condition allows us to concentrate on situations in which the
scattering is governed by the anisotropic part of the dielectric
properties.  This distinguishes our case from other studies of 
scattering by spheres, in which the isotropic optical contrast
dominates. However, there is also a motivation for this
hypothesis in terms of liquid crystal  device physics, and we shall 
discuss this at greater length in a subsequent paper. However, the result 
of the hypothesis is that the scattering by our model spheres
disappears in the limit of zero anisotropy.

\section{\T--matrix approach in isotropic medium}
\label{sec:t-matrix-approach}

\subsection{\T--matrix ansatz}
\label{subsec:t-matrix-ansatz}

In this subsection we remind the reader about the relationship between
Maxwell's equations in the region of a scatterer and the formulation
of scattering properties in terms of the
\T--matrix~\cite{New,Ishim}. Our formulation is slightly
non-standard. Some technical details, which can be omitted at first
reading, have been relegated to the appendices.

We shall need to write the  Maxwell equations 
for a harmonic electromagnetic wave
(time--dependent factor is $\exp\{-i\omega t\}$) in the
form:
\begin{subequations}
  \label{eq:maxwell}
\begin{align}
-i n_i[\mu_ik_i]^{-1}\,
\bs{\nabla}\times\vc{E}&=\vc{H}\, ,
\label{eq:maxwell1}\\
i \mu_i[n_i k_i]^{-1}\,
\bs{\nabla}\times\vc{H}&=
\vc{E}
+u_i\bigl(\uvc{n}\cdot\vc{E}\bigr)\uvc{n}\, .
\label{eq:maxwell2}
\end{align}
\end{subequations}
where $n_i=\sqrt{\epsilon_i\mu_i}$ are refractive indexes
for the regions, where $R_2<r<R_1$ ($i=1$)
and $r<R_2$ ($i=2$); $k_i=n_ik_{vac}$ 
($k_{vac}=\omega/c=2\pi/\lambda$ is the free--space wavenumber).
We define the \emph{anisotropy parameter} as 
$u_1=\Delta\epsilon_1/\epsilon_1$ (in the annular layer). Then inside
the isotropic core  $u_2=0$.
Finally, for brevity, in the region outside the scatterer $r>R_1$, 
the index will be suppressed, giving
$k\equiv nk_{vac}$ and $u=0$.

The electromagnetic field can 
always be expanded
using the vector spherical harmonic basis,  
$\vc{Y}_{j+\delta\, j\, m}(\phi,\theta)\equiv
\vc{Y}_{j+\delta\, j\, m}(\uvc{r})$ 
($\delta=0,\pm 1$)~\cite{Bie},
as follows:
\begin{widetext}
\begin{subequations}
\label{eq:spher}
\begin{align}
& \vc{E}=\sum_{jm}\vc{E}_{jm}=\sum_{jm}\left[
       p^{(0)}_{jm}(r) \vc{Y}^{(0)}_{jm}(\uvc{r})+
       p^{(e)}_{jm}(r) \vc{Y}^{(e)}_{jm}(\uvc{r})+
       p^{(m)}_{jm}(r) \vc{Y}^{(m)}_{jm}(\uvc{r})\right]\:, 
\label{eq:e_spher}\\
& \vc{H}=\sum_{jm}\vc{H}_{jm}=\sum_{jm}\left[
       q^{(0)}_{jm}(r) \vc{Y}^{(0)}_{jm}(\uvc{r})+
       q^{(e)}_{jm}(r) \vc{Y}^{(e)}_{jm}(\uvc{r})+
       q^{(m)}_{jm}(r) \vc{Y}^{(m)}_{jm}(\uvc{r})\right]\:,
\label{eq:h_spher}
\end{align}
\end{subequations}
\end{widetext}
where
$\vc{Y}^{(m)}_{jm}=\vc{Y}_{j\,j\,m}$ and 
$\vc{Y}^{(e)}_{jm}=[j/(2j+1)]^{1/2}\vc{Y}_{j+1\,j\,m}+
[(j+1)/(2j+1)]^{1/2}\vc{Y}_{j-1\,j\,m}$ are electric and 
magnetic harmonics respectively, and 
$\vc{Y}^{(0)}_{jm}=[j/(2j+1)]^{1/2}\vc{Y}_{j-1\,j\,m}-
[(j+1)/(2j+1)]^{1/2}\vc{Y}_{j+1\,j\,m}$ are  longitudinal harmonics
(a number of relations for the vector spherical harmonics
used throughout this paper
are considered in Appendix~\ref{sec:vect-spher-harm}).
The electric field is now completely described by the coefficients
$\{p_{jm}^{(\alpha)}(r)\}$
and similarly the magnetic field is now described by
$\{q_{jm}^{(\alpha)}(r)\}$ with $\alpha=\{o,e,m\}$.

In order to find the coefficient functions we  use separation of
variables.  This implies that the expansions~\eqref{eq:spher} must be
inserted into Maxwell's equations~\eqref{eq:maxwell}. The coefficient
functions then can be derived by solving the resulting system of
equations.  In the simplest case of  an isotropic medium the coefficient
functions  
can be expressed in terms of spherical Bessel functions,
$j_j(x)=[\pi/(2x)]^{1/2} J_{j+1/2}(x)$, and spherical Hankel
functions~\cite{Abr}, $h_j^{(1)}(x)=[\pi/(2x)]^{1/2}
H_{j+1/2}^{(1)}(x)$, and their derivatives as follows
\begin{widetext}
\begin{subequations}
\label{eq:sep}
\begin{align}
& \vc{E}_{jm}=
\alpha_{jm}\vc{M}^{(m)}_{jm}(\rho,\uvc{r})+
\beta_{jm}\vc{\tilde{M}}^{(m)}_{jm}(\rho,\uvc{r})-\frac{\mu}{n}
\left(
\tilde\alpha_{jm}\vc{M}^{(e)}_{jm}(\rho,\uvc{r})+
\tilde\beta_{jm}\vc{\tilde{M}}^{(e)}_{jm}(\rho,\uvc{r}) \right)\, ,
\label{eq:sep1}\\
& \vc{H}_{jm}=
\tilde\alpha_{jm}\vc{M}^{(m)}_{jm}(\rho,\uvc{r})+
\tilde\beta_{jm}\vc{\tilde{M}}^{(m)}_{jm}(\rho,\uvc{r})+\frac{n}{\mu}
\left(
\alpha_{jm}\vc{M}^{(e)}_{jm}(\rho,\uvc{r})+
\beta_{jm}\vc{\tilde{M}}^{(e)}_{jm}(\rho,\uvc{r}) \right)\, ,
\label{eq:sep2}
\end{align} 
\end{subequations}
\end{widetext}
where $\alpha_{jm}$, $\tilde\alpha_{jm}$, $\beta_{jm}$ and
$\tilde\beta_{jm}$ are integration constants; the vector functions
$\vc{M}_{jm}^{(\alpha)}$ and $\vc{\tilde{M}}_{jm}^{(\alpha)}$ are
given by
\begin{widetext}
\begin{subequations}
  \label{eq:iso}
\begin{align}
&  \vc{M}^{(m)}_{jm}(\rho,\uvc{r})
=j_j(\rho)\vc{Y}^{(m)}_{jm}(\uvc{r})\, ,\notag\\
&  \vc{M}^{(e)}_{jm}(\rho,\uvc{r})
=Dj_j(\rho)\vc{Y}^{(e)}_{jm}(\uvc{r})+
\frac{\sqrt{j(j+1)}}{\rho}\, j_j(\rho)\vc{Y}^{(o)}_{jm}(\uvc{r})\, ,
\label{eq:iso1}\\
&  \vc{\tilde{M}}^{(m)}_{jm}(\rho,\uvc{r})
=h^{(1)}_j(\rho)\vc{Y}^{(m)}_{jm}(\uvc{r})\, ,\notag\\
&  \vc{\tilde{M}}^{(e)}_{jm}(\rho,\uvc{r})
=Dh^{(1)}_j(\rho)\vc{Y}^{(e)}_{jm}(\uvc{r})+
\frac{\sqrt{j(j+1)}}{\rho}\, h^{(1)}_j(\rho)\vc{Y}^{(o)}_{jm}(\uvc{r})\, ,
\label{eq:iso2}
\end{align}
\end{subequations}
\end{widetext}
where $Df(x)\equiv x^{-1}\dfrac{\dd}{\dd x}(xf(x))$ and
$\rho\equiv kr$.

There are two cases of Eq.~\eqref{eq:sep1} that are of particular
interest. They correspond to the incoming incident wave,
$\{\vc{E}_{inc},\vc{H}_{inc}\}$, and the outgoing scattered wave,
$\{\vc{E}_{sca},\vc{H}_{sca}\}$:
\begin{widetext}
\begin{gather}
\vc{E}_{jm}^{(inc)}=
\alpha_{jm}^{(inc)}\,\vc{M}^{(m)}_{jm}(\rho,\uvc{r})
-\frac{\mu}{n}\,
\tilde\alpha_{jm}^{(inc)}\,\vc{M}^{(e)}_{jm}(\rho,\uvc{r})\notag\, ,\\
\vc{H}_{jm}^{(inc)}=
\tilde\alpha_{jm}^{(inc)}\,\vc{M}^{(m)}_{jm}(\rho,\uvc{r})
+\frac{n}{\mu}\,
\alpha_{jm}^{(inc)}\,\vc{M}^{(e)}_{jm}(\rho,\uvc{r})\, ,
\label{eq:inc}
\end{gather}
\begin{gather}
\vc{E}_{jm}^{(sca)}=
\beta_{jm}^{(sca)}\,\vc{\tilde{M}}^{(m)}_{jm}(\rho,\uvc{r})
-\frac{\mu}{n}\,
\tilde\beta_{jm}^{(sca)}\,\vc{\tilde{M}}^{(e)}_{jm}(\rho,\uvc{r})
\notag\, .\\
\vc{H}_{jm}^{(sca)}=
\tilde\beta_{jm}^{(sca)}\,\vc{\tilde{M}}^{(m)}_{jm}(\rho,\uvc{r})
+\frac{n}{\mu}\,
\beta_{jm}^{(sca)}\,\vc{\tilde{M}}^{(e)}_{jm}(\rho,\uvc{r})
\, .
\label{eq:sca}
\end{gather}
\end{widetext}
Now the incoming incident wave is characterised by amplitudes
$\alpha_{jm}^{(inc)}$, $\tilde{\alpha}_{jm}^{(inc)}$ and the scattered
outgoing waves are similarly characterised by amplitudes
$\beta_{jm}^{(sca)}$, $\tilde{\beta}_{jm}^{(sca)}$.  Our task is now
to relate $\{\alpha\}$ and $\{\beta\}$.

In this regime, a transverse plane wave incident in the direction is
specified by an unit vector $\uvc{k}_{inc}$, with
\begin{widetext}
\begin{equation}
  \label{eq:polar_inc}
\vc{E}_{inc}=\vc{E}^{(inc)}\exp(i\,\vc{k}_{inc}\cdot\vc{r})\, ,
\quad
\vc{E}^{(inc)}=\sum_{\nu=\pm 1}
E_{\nu}^{(inc)}\vc{e}_{\nu}(\uvc{k}_{inc})\, ,
\qquad
\vc{k}_{inc}=k\uvc{k}_{inc}\, .
\end{equation}
\end{widetext}
We show in~\eqref{eq:B.5} the coefficients $\{\alpha\}$ of the
expansion~\eqref{eq:inc} takes 
the form:
\begin{widetext}
\begin{align}
  \label{eq:coef_inc}
\alpha_{jm}^{(inc)}&= i
\sum_{\nu=\pm 1} i^{j+1}[2\pi(2j+1)]^{1/2} 
D_{m \nu}^{j}(\uvc{k}_{inc})\nu 
E_{\nu}^{(inc)}\, ,\notag
\\
\tilde{\alpha}_{jm}^{(inc)}&= n/\mu
\sum_{\nu=\pm 1} i^{j+1}[2\pi(2j+1)]^{1/2} 
D_{m \nu}^{j}(\uvc{k}_{inc})
E_{\nu}^{(inc)}\, , 
\end{align}
\end{widetext}
where $D_{mm'}^j$ is the Wigner $D$-function~\cite{Bie,Gel} and the
basis vectors $\vc{e}_{\pm 1}(\uvc{k}_{inc})$ are perpendicular to
$\uvc{k}_{inc}$ and defined by Eq.~\eqref{eq:vec_k}.

Thus outside the scatterer the electromagnetic field is a sum of the
transverse plane wave incident in the direction specified by an unit
vector $\uvc{k}_{inc}$ ($\beta_{jm}^{(inc)}=
\tilde{\beta}_{jm}^{(inc)}=0$) and the outgoing wave with
$\alpha_{jm}^{(sca)}=\tilde{\alpha}_{jm}^{(sca)}=0$ as required by the
Sommerfeld radiation condition.  

So long as the scattering problem is
linear, the coefficients $\beta_{jm}^{(sca)}$ and
$\tilde{\beta}_{jm}^{(sca)}$ can be written as linear combinations of
$\alpha_{jm}^{(inc)}$ and $\tilde{\alpha}_{jm}^{(inc)}$:
\begin{widetext}
\begin{gather}
  \beta_{jm}^{(sca)}=\sum_{j',m'}\left[\,
T_{jm,\,j'm'}^{\,11}\, \alpha_{j'm'}^{(inc)}+\frac{\mu}{n}\, 
T_{jm,\,j'm'}^{\,12}\,\tilde{\alpha}_{j'm'}^{(inc)}
\,\right],\notag\\
\tilde{\beta}_{jm}^{(sca)}=\sum_{j',m'}\left[\,
\frac{n}{\mu}\,T_{jm,\,j'm'}^{\,21}\, \alpha_{j'm'}^{(inc)}+ 
T_{jm,\,j'm'}^{\,22}\,\tilde{\alpha}_{j'm'}^{(inc)}
\,\right]\, .
  \label{eq:matr}
\end{gather}
\end{widetext}
These formulae define the elements of the \T--matrix in the most general case.

In general, the outgoing wave with angular momentum index $j$ arises
from ingoing waves of all other indices $j'$. In such cases we say
that the scattering process mixes angular momenta~\cite{Mis:1996}.
The light scattering from
the uniformly anisotropic scatterer, depicted in Fig.~\ref{fig:str}b,
provides an example of such a scattering process.
In this case the cylindrical symmetry of the optical axis distribution causes  
the \T-matrix to be diagonal over azimuthal indices $m$ and $m'$: 
$T_{jm,\,j'm'}^{\,nn'}=\delta_{mm'}\,T_{jj';\,m}^{\,nn'}$. Then we can
conveniently rewrite  the relation~\eqref{eq:matr} using   matrix
notations:
\begin{widetext}
\begin{equation}
\begin{pmatrix}
\beta_{jm}^{(sca)}\\
{\mu}/n\, 
\tilde{\beta}_{jm}^{(sca)}
\end{pmatrix}
=\sum_{j'} \left[\vc{T}\right]_{jj';\,m}
\begin{pmatrix}
\alpha_{j'm}^{(inc)}\\
{\mu}/n\, 
\tilde{\alpha}_{j'm}^{(inc)}
\end{pmatrix}\, ,\quad
\left[\vc{T}\right]_{jj';\,m}=
\begin{pmatrix}
T_{jj';\,m}^{\,11}& T_{jj';\,m}^{\,12}\\
T_{jj';\,m}^{\,21}& T_{jj';\,m}^{\,22}\\
\end{pmatrix}\, .
  \label{eq:matr_uni}
\end{equation}
\end{widetext}

In simpler scattering processes, by contrast, such angular momentum
mixing does not take place. Many quantum scattering processes and
classical Mie scattering belong to this category.  
It is seen from Fig.~\ref{fig:str}a  that 
radial anisotropy keeps intact spherical symmetry of the scatterer.
The radially anisotropic annular layer thus exemplifies a scatterer
that does not mix angular momenta. 
The \T--matrix of a spherically symmetric scatterer is diagonal
over the angular momenta and the azimuthal numbers:
$T_{jj',mm'}^{nn'}=\delta_{jj'}\delta_{mm'} T_{j}^{nn'}$.

\subsection{Scattering Amplitude Matrix}
\label{subsec:scatt-ampl-matr}

We have seen the relation between the scattered wave~\eqref{eq:sca} and the
incident plane wave~\eqref{eq:inc} is linear.  In the far field region
($\rho\gg 1$), where the asymptotic behaviour of the spherical Bessel
and Hankel functions is known~\cite{Abr}:
\begin{widetext}
\begin{equation}
  \label{eq:asymp}
i^{j+1}h_j^{(1)}(\rho), i^{j}Dh_j^{(1)}(\rho)\sim\exp(i\rho)/\rho\, .
\end{equation}
\end{widetext}
The scattering amplitude matrix
$\vc{A}(\uvc{k}_{sca},\uvc{k}_{inc})$, which relates
$\vc{E}_{sca}$ and the polarisation vector of the incident wave
$\vc{E}^{(inc)}$ is defined in the following 
way~\cite{New,Ishim,Mis:1996}:
\begin{widetext}
\begin{equation}
\label{eq:ampl_def}
E_{\nu}^{(sca)}\equiv
(\vc{e}_{\nu}^{\,*}(\uvc{k}_{sca}),
\vc{E}_{sca})=
\rho^{-1}\exp(i\rho)\sum_{\nu'=\pm 1} 
\vc{A}_{\nu\nu'}(\uvc{k}_{sca},\uvc{k}_{inc}) 
E_{\nu'}^{(inc)}\, ,\quad \nu=\pm 1\,
\end{equation}
\end{widetext}
where an asterisk indicates complex conjugation,
$\uvc{k}_{sca}=\uvc{r}$ and
$\uvc{e}_{\pm 1}(\uvc{k}_{sca})=\mp 
(\,\ubs{\vartheta}\pm i\ubs{\varphi}\,)/\sqrt{2}$.

Eqs.~\eqref{eq:sca},~\eqref{eq:asymp} and the vector spherical
harmonic relations Eqs.~\eqref{eq:A.y2d} can now be combined to yield
the expression for the scattering amplitude matrix in terms of the
$T$--matrix:
\begin{widetext}
\begin{align}
\vc{A}_{\nu\nu'}(\uvc{k}_{sca},\uvc{k}_{inc})&=-\frac{i}{2}\,
\sum_{jm}\sum_{j'm'}\,[(2j+1)(2j'+1)]^{1/2}
D_{m\nu}^{j\,*}(\uvc{k}_{sca})D_{m'\nu'}^{j'}(\uvc{k}_{inc})\cdot
\notag\\
&\cdot\left[\,
\nu\nu'\, T_{jm,\,j'm'}^{\,11}-i\nu\, T_{jm,\,j'm'}^{\,12}+
i\nu'\, T_{jm,\,j'm'}^{\,21}+T_{jm,\,j'm'}^{\,22}\,
\right]\, .
\label{eq:ampl}  
\end{align}
\end{widetext}

For a spherically symmetric scatterer, 
this result can be expressed in the simplified form:
\begin{widetext}
\begin{subequations}
  \label{eq:ampl_spher_sym}
\begin{align}
\vc{A}_{\nu\nu'}(\uvc{k}_{sca},\uvc{k}_{inc})&=
\sum_j \vc{A}_{\nu\nu'}^{j}(\uvc{k}_{sca},\uvc{k}_{inc})=\notag\\
=&-i\sum_{j} (j+1/2)\tilde{D}_{\nu\nu'}^{j}
(\uvc{k}_{sca},\uvc{k}_{inc})\,
\left[\,
\nu\nu'\, T_j^{\,11}-i\nu\, T_j^{\,12}+i\nu'\, T_j^{\,21}+T_j^{\,22}\,
\right]\, ,
\label{eq:ampl_spher_a}\\
& \tilde{D}_{\nu\nu'}^{j}(\uvc{k}_{sca},\uvc{k}_{inc})=
\sum_m D_{m\nu}^{j\,*}(\uvc{k}_{sca})
D_{m\nu'}^{j}(\uvc{k}_{inc})\, .
\label{eq:ampl_spher_b}
\end{align}
\end{subequations}
\end{widetext}
Eq.~\eqref{eq:ampl_spher_b} shows that
the scattering amplitude matrix~\eqref{eq:ampl_spher_a} 
depends only on the angle between $\uvc{k}_{inc}$ and $\uvc{k}_{inc}$.

All scattering properties of the system can be computed
from the elements of the scattering amplitude matrix.
In Eq.~\eqref{eq:ampl} we see that this can be defined in terms of the
elements of the $T$--matrix defined in Eq.~\eqref{eq:matr}. Thus
computation of these matrix elements is of crucial importance.

\subsection{Scattering Efficiency}
\label{subsec:scatt-effic}

In order to find the total scattering cross section $C_{sca}$ we need
to calculate the flux of Poynting vector of the scattered wave
$\vc{S}^{(sca)}=c/(8\pi)\Re(\vc{E}_{sca}\times\vc{H}_{sca}^{\,*})$
through a sphere of sufficiently large radius and divide the result
by $|\vc{S}^{(inc)}|$.

From Eqs.~\eqref{eq:sca} and~\eqref{eq:asymp} the asymptotic behaviour of
the scattered outgoing wave in the far-field region is given by
\begin{widetext}
\begin{subequations}
\label{eq:sca_asymp}
\begin{align}
i^{j+1}\vc{E}_{jm}^{(sca)}\sim 
\left[\frac{2j+1}{8\pi}\right]^{1/2}&\frac{\exp(i\rho)}{\rho}\,\biggl\{
\left(
\beta_{jm}^{(sca)}-i\,\mu n^{-1}\tilde{\beta}_{jm}^{(sca)}
\right)
D_{m,\,-1}^{j\,*}(\uvc{k}_{sca})\,\vc{e}_{-1}(\uvc{k}_{sca})-\notag\\
&-\left(
\beta_{jm}^{(sca)}+i\,\mu n^{-1}\tilde{\beta}_{jm}^{(sca)}
\right)
D_{m,\,1}^{j\,*}(\uvc{k}_{sca})\,\vc{e}_{+1}(\uvc{k}_{sca})
\biggr\}\, ,
  \label{eq:e_asymp}\\
i^{j+1}\vc{H}_{jm}^{(sca)}\sim 
\left[\frac{2j+1}{8\pi}\right]^{1/2}&\frac{\exp(i\rho)}{\rho}\,\biggl\{
\left(
\tilde{\beta}_{jm}^{(sca)}+i\,\mu^{-1} n\beta_{jm}^{(sca)}
\right)
D_{m,\,-1}^{j\,*}(\uvc{k}_{sca})\,\vc{e}_{-1}(\uvc{k}_{sca})-\notag\\
&-\left(
\tilde{\beta}_{jm}^{(sca)}-i\,\mu^{-1} n \beta_{jm}^{(sca)}
\right)
D_{m,\,1}^{j\,*}(\uvc{k}_{sca})\,\vc{e}_{+1}(\uvc{k}_{sca})
\biggr\}\, ,
  \label{eq:h_asymp}
\end{align}
\end{subequations}
\end{widetext}
where the relations between the vector spherical functions
and Wigner $D$-functions~\eqref{eq:A.y2d} have been used.

The asymptotic relation~\eqref{eq:e_asymp} give
the relations~\eqref{eq:ampl_def} and~\eqref{eq:ampl}
that define the scattering matrix $\vc{A}$. Similarly,
from Eq.~\eqref{eq:h_asymp} we have the following relation
for magnetic field of the scattered wave in the far field region:
\begin{widetext}
\begin{equation}
\label{eq:ampl_def_h}
\mu/n\,H_{\nu}^{(sca)}\equiv \mu/n\,
(\vc{e}_{\nu}^{\,*}(\uvc{k}_{sca}),
\vc{H}_{sca})= -i
\rho^{-1}\exp(i\rho)\sum_{\nu'=\pm 1} 
\nu\,\vc{A}_{\nu\nu'}(\uvc{k}_{sca},\uvc{k}_{inc}) 
E_{\nu'}^{(inc)}\, ,\quad \nu=\pm 1\,.
\end{equation}
\end{widetext}

From Eqs.~\eqref{eq:ampl_def},~\eqref{eq:ampl_def_h} and~\eqref{eq:A.3}
we can now readily express the Poynting vector in terms of the
scattering matrix:
\begin{widetext}
\begin{equation}
  \label{eq:poynt_vec}
  \mu/n\,\vc{E}_{sca}\times\vc{H}_{sca}^{\,*}=\uvc{k}_{sca}\,\rho^{-2}
\sum_{\nu,\,\nu'=\pm 1}E_{\nu}^{(inc)}\, 
\left[\vc{A}\cdot\vc{A}^{\dagger}\right]_{\nu\nu'}\,
E_{\nu'}^{(inc)\,*}\, ,
\end{equation}
\end{widetext}
where the superscript $\dagger$ indicates Hermitian conjugation.

Using these expressions and the orthogonality relation~\eqref{eq:A.9},
we can immediately deduce the result for the total scattering
cross-section:
\begin{widetext}
\begin{equation}
  \label{eq:cross}
  C_{sca}=k^{-2} I_{inc}^{-1}\sum_{jm}
\left[\,
\bigl|\beta_{jm}^{(sca)}\bigr|^2+
\bigl|\mu n^{-1}\tilde{\beta}_{jm}^{(sca)}\bigr|^2
\,\right]\, ,
\end{equation}
\end{widetext}
where 
$\displaystyle
I_{inc}=\sum_{\nu=\pm 1}\bigl|E_{\nu}^{(inc)}\bigr|^2\,
$.

Integrating 
the product of matrices that enter Eq.~\eqref{eq:poynt_vec}
over the angles of scattered wave gives
\begin{widetext}
\begin{align}
&
\langle\,
\left[\vc{A}\cdot\vc{A}^{\dagger}\right]_{\nu\nu'}
\,\rangle_{sca}= 2\pi
\sum_{jm}\sum_{j'm'}\,
[(2j+1)(2j'+1)]^{1/2}\,
D_{m\nu}^{j\,*}(\uvc{k}_{inc})\,
D_{m'\nu'}^{j'}(\uvc{k}_{inc})\,\cdot\notag\\
&
\cdot\left[
(i\nu\,\vc{T}^{11}+\vc{T}^{12})^{\dag}\cdot
(i\nu'\,\vc{T}^{11}+\vc{T}^{12})+
(i\nu\,\vc{T}^{21}+\vc{T}^{22})^{\dag}\cdot
(i\nu'\,\vc{T}^{21}+\vc{T}^{22})
\right]_{jm,\,j'm'}\, ,
\label{eq:aa-mix}
\end{align}
\end{widetext}
where
$\displaystyle
\langle\,f\,\rangle_{sca}\equiv\int_0^{2\pi}\dd\phi_{sca}
\int_0^{\pi}\sin\theta_{sca}\,\dd\theta_{sca}\,f$.

So, we have 
the matrix $\langle\,\vc{A}\cdot\vc{A}^{\dagger}\,\rangle_{sca}$
which is, in general, non-diagonal and depend on incident wave angles.
For a spherically symmetric scatterer this matrix is
diagonal and independent of $\uvc{k}_{inc}$.
In this case the cross-section~\eqref{eq:cross} can be written
in the following form:
\begin{widetext}
\begin{equation}
  C_{sca}=\frac{1}{2k^{2}}
\Tr\,\langle\vc{A}\cdot\vc{A}^{\dagger}\rangle_{sca}\, .
  \label{eq:cross_spher}
\end{equation}
\end{widetext}
Note that for unpolarised incident light this relation holds
even if a scatterer is not spherically symmetric.
 
We can now relate the scattering cross-section and the
elements of the \T-matrix.  
For brevity, we restrict ourselves to the
spherically symmetric case, considered at the end of the previous
subsection.  More precisely, we consider the scattering efficiency,
$Q$, that is the ratio of $C_{sca}$ and area of the composite
particle, $S=\pi R_1^2$.  
For diagonal \T-matrix, Eqs.~\eqref{eq:cross_spher} and~\eqref{eq:aa-mix} 
give the known result~\cite{Mis:1996}:
\begin{widetext}
\begin{equation}
  \label{eq:effic_spher}
  Q=\frac{C_{sca}}{S}=
\frac{2}{k^2 R_1^2}\sum_{j=1}^{\infty}\sum_{m,n=1}^2(2j+1)\,
\left|T_j^{mn}\right|^2\, .
\end{equation}
\end{widetext}

In order to characterise angular distribution of scattered light
intensity let us suppose that the incident wave is linearly polarised
and is propagating along the $z$-axis, $\uvc{k}_{inc}=\uvc{z}$.
The wave vectors $\uvc{k}_{inc}$ and $\uvc{k}_{sca}$ define the
scattering plane. From Eqs.~\eqref{eq:ampl_def}
and~\eqref{eq:ampl_spher_sym} we can find the amplitudes of the scattered
wave components that are parallel 
($E_{x}^{(sca)}$) and normal ($E_{y}^{(sca)}$) to the scattering
plane:
\begin{widetext}
\begin{subequations}
  \label{eq:par-per}
\begin{align}
&\lvert E_{x}^{(sca)}\rvert^2(\theta_{sca})=I_{inc}\, 
i_{\parallel}(\theta_{sca})\, \cos^2\psi\, ,
\label{eq:tan}\\
&\lvert E_{y}^{(sca)}\rvert^2(\theta_{sca})=I_{inc}\, 
i_{\perp}(\theta_{sca})\, \sin^2\psi\, ,
\label{eq:perp}\\
&i_{\parallel}=\lvert i_1+i_{-1} \rvert^2\, ,\quad
i_{\perp}=\lvert i_1-i_{-1} \rvert^2\, ,
\label{eq:i-par-per}\\
& i_{\nu}(\theta_{sca})=
(kR_1)^{-1}\sum_j (j+1/2)\, d_{1\,\nu}^{\,j}(\theta_{sca})\,
[\,T_j^{11}+\nu T_j^{22}\,]\, ,
\end{align}  
\end{subequations}
\end{widetext}
where $\psi$ is the angle between the polarisation vector of the
incident wave and the scattering plane.

In subsequent sections
we shall use the intensity $i_{sca}(\theta_{sca})$
\begin{widetext}
\begin{equation}
  \label{eq:ang-int}
i_{sca}(\theta_{sca})=
\left(
  i_{\parallel}(\theta_{sca})+i_{\perp}(\theta_{sca})
\right) /2
\end{equation}
\end{widetext}
and the factor characterising the degree of depolarisation
(depolarisation factor) $P_{dep}(\theta_{sca})$
\begin{widetext}
\begin{equation}
  \label{eq:pol-fact}
P_{dep}(\theta_{sca})=1-P(\theta_{sca}),\quad
  P(\theta_{sca})=
\frac{\lvert i_{\parallel}(\theta_{sca})-i_{\perp}(\theta_{sca})
  \rvert}{i_{\parallel}(\theta_{sca})+i_{\perp}(\theta_{sca})}
\end{equation}
\end{widetext}
as quantities characterising angular distribution and polarisation 
of the scattered wave~\cite{Born}. 
Note that averaging $i_{sca}(\theta_{sca})$
over the scattering angle gives the scattering efficiency:
$$
Q=\int_0^\pi i_{sca}(\theta_{sca})\sin\theta_{sca}\,\dd\theta_{sca}\, .
$$

\section{Scattering from radially anisotropic layer}
\label{sec:scatt-from-radi}

In Sec.~\ref{subsec:t-matrix-ansatz} we started from the general
expansion for electromagnetic field over the vector spherical
harmonics~\eqref{eq:spher}. Then the fields in isotropic medium
were expressed in terms of the modes, $\vc{M}_{jm}^{(\,\alpha)}$  and
$\tilde{\vc{M}}_{jm}^{(\,\alpha)}$ (see Eq.~\eqref{eq:iso}).
This expression is known as the \T-matrix ansatz~\cite{Mis:1996,Mish}.

We shall write down the results for electromagnetic field
within the radially anisotropic layer
as they are given in~\cite{Kis:2000:opt}.
These can be written in the form similar to the \T-matrix ansatz:
\begin{widetext}
\begin{subequations}
\label{eq:anis}
\begin{align}
& \vc{E}_{jm}=
\alpha_{jm}\vc{P}^{(m)}_{jm}(\rho,\uvc{r})+
\beta_{jm}\vc{\tilde{P}}^{(m)}_{jm}(\rho,\uvc{r})-\frac{\mu}{n}
\left(
\tilde\alpha_{jm}\vc{P}^{(e)}_{jm}(\rho,\uvc{r})+
\tilde\beta_{jm}\vc{\tilde{P}}^{(e)}_{jm}(\rho,\uvc{r}) \right)\, ,
\label{eq:anis_e}\\
& \vc{H}_{jm}=
\tilde\alpha_{jm}\vc{Q}^{(m)}_{jm}(\rho,\uvc{r})+
\tilde\beta_{jm}\vc{\tilde{Q}}^{(m)}_{jm}(\rho,\uvc{r})+\frac{n}{\mu}
\left(
\alpha_{jm}\vc{Q}^{(e)}_{jm}(\rho,\uvc{r})+
\beta_{jm}\vc{\tilde{Q}}^{(e)}_{jm}(\rho,\uvc{r}) \right)\, .
\label{eq:anis_h}
\end{align} 
\end{subequations}
For radial anisotropy the modes that enter Eqs.~\eqref{eq:anis}
are given by
\begin{subequations}
\label{eq:rad_bas}
\begin{align}
\vc{Q}^{(m)}_{jm}=j_{\tilde{j}}(\rho_1)\,\vc{Y}^{(m)}_{jm}(\uvc{r})
\, ,\quad
\vc{P}^{(e)}_{jm}=
Dj_{\tilde{j}}(\rho_1)\,\vc{Y}^{(e)}_{jm}(\uvc{r})
+[\tilde{j}(\tilde{j}+1)]^{1/2}\,\rho_1^{-1}\,
j_{\tilde{j}}(\rho_1)\,\vc{Y}^{(o)}_{jm}(\uvc{r})\, ,
\label{eq:rad_bas1}\\
\vc{P}^{(m)}_{jm}=\vc{M}^{(m)}_{jm}(\rho_1,\uvc{r})\, ,\quad
\vc{Q}^{(e)}_{jm}=\vc{M}^{(e)}_{jm}(\rho_1,\uvc{r})\, ,\quad
\tilde{j}(\tilde{j}+1)=j(j+1)/(1+u_1)\, .
\label{eq:rad_bas2}
\end{align}
\end{subequations}
\end{widetext}

In the next section we shall find that the fields in
uniformly anisotropic medium can also be written in the 
form~\eqref{eq:anis}. However, the expressions for the normal modes
in this case will differ from those of Eqs.~\eqref{eq:rad_bas}.
In both cases  the modes represent solutions of Maxwell's equations
and  turn into the isotropic medium modes~\eqref{eq:iso}
in the limit of small anisotropy parameter, $u\to 0$.

The radial anisotropy harmonics~\eqref{eq:anis} do not involve 
angular momentum mixing and as such they only have contributions proportional
to vector spherical harmonics with angular momentum numbers
given $j$ and $m$. The important difference between the two cases lies
in the fact  
that for the uniformly anisotropic layer this is no longer true. 

The fields inside the isotropic core of the particle 
similarly also involve no angular momentum mixing:
\begin{widetext}
\begin{subequations}
\label{eq:core}
\begin{align}
& \vc{E}_{jm}^{(c)}=
\alpha_{jm}^{(c)}\,\vc{M}^{(m)}_{jm}(\rho_2,\uvc{r})
-\frac{\mu_2}{n_2}\,
\tilde\alpha_{jm}^{(c)}\,\vc{M}^{(e)}_{jm}(\rho_2,\uvc{r})\, ,
\label{eq:core_e}\\
& \vc{H}_{jm}^{(c)}=
\tilde\alpha_{jm}^{(c)}\,\vc{M}^{(m)}_{jm}(\rho_2,\uvc{r})+\frac{n_2}{\mu_2}\,
\alpha_{jm}^{(c)}\,\vc{M}^{(e)}_{jm}(\rho_2,\uvc{r})\, ,
\label{eq:core_h}
\end{align} 
\end{subequations}
\end{widetext}
where $\rho_i\equiv k_i r$.

We now pass on to the calculation of the scattering cross-section
using the \T--matrix method. 
The formulae presented in this subsection form a key element in the
input to this calculation.

\subsection{\T--matrix: radial anisotropy}
\label{subsec:t-matr-scatt}
 
In order to calculate the elements of \T--matrix, we need to use
continuity of the tangential components of the electric and magnetic
fields as boundary conditions at $r=R_2$ and $r=R_1$.
Equivalently, the functions 
$p_{jm}^{(\alpha)}\equiv
\langle\vc{Y}_{jm}^{(\alpha)\,*}\cdot\vc{E}\rangle_{\uvc{r}}$,
and
$q_{jm}^{(e)}\equiv
\langle\vc{Y}_{jm}^{(\alpha)\,*}\cdot\vc{H}\rangle_{\uvc{r}}$,
$\alpha=m, e$,
must be continuous at each boundary:
\begin{widetext}
\begin{equation}
p_{jm}^{(\alpha)}(R_i+0)=p_{jm}^{(\alpha)}(R_i-0),\quad
q_{jm}^{(\alpha)}(R_i+0)=q_{jm}^{(\alpha)}(R_i-0),\quad
i=1, 2\,;\,\alpha=m, e\,.
\label{eq:cont}
\end{equation}
\end{widetext}
For the radially anisotropic distribution~\eqref{eq:rad} different angular
momentum are decoupled. We have shown elsewhere that the \T--matrix
can be computed in the 
closed form~\cite{Kis:2000:opt}, and here we give more details of this
calculation.  

Expressions for $\{p_{jm}^{(\alpha)},q_{jm}^{(\alpha)}\}$ in the
ambient medium have been given implicitly in Eqs.~\eqref{eq:inc}
and~\eqref{eq:sca}.  
Similar expressions in the annular layer
have been given by Eqs.~\eqref{eq:anis} and~\eqref{eq:rad_bas}. Finally, 
the analogous expressions inside the isotropic core have been given
in Eq.~\eqref{eq:core}. 

In order to determine the \T-matrix we shall insert these expressions 
into the boundary conditions~\eqref{eq:cont}. This will yield
a system of eight linear algebraic equations for the ten quantities:
$\alpha_{jm}^{(inc)}$, $\tilde{\alpha}_{jm}^{(inc)}$, 
$\beta_{jm}^{(sca)}$, $\tilde{\beta}_{jm}^{(sca)}$, 
$\alpha_{jm}$, $\beta_{jm}$,
$\tilde{\alpha}_{jm}$, $\tilde{\beta}_{jm}$, 
$\alpha_{jm}^{(c)}$ and $\tilde{\alpha}_{jm}^{(c)}$. 
After eliminating all internal variables we 
shall be left with two equations in the four unknowns: 
$\alpha_{jm}^{(inc)}$, $\tilde{\alpha}_{jm}^{(inc)}$,
$\beta_{jm}^{(sca)}$, $\tilde{\beta}_{jm}^{(sca)}$. 
These equations will define the \T-matrix.  

In order to carry out this procedure efficiently 
we combine Eqs.~\eqref{eq:spher} and ~\eqref{eq:anis}
by using the compact matrix notation
for the components inside the anisotropic layer:
\begin{widetext}
\begin{equation}
  \label{eq:matr_not}
\begin{pmatrix}
p_{jm}^{(m)}(r)\\
q_{jm}^{(e)}(r)\\
q_{jm}^{(m)}(r)\\
p_{jm}^{(e)}(r)
\end{pmatrix}
=\vc{R}(r)\begin{pmatrix}
\alpha_{jm}\\
\beta_{jm}\\
\tilde{\alpha}_{jm}\\
\tilde{\beta}_{jm}
\end{pmatrix}\, ,
\end{equation}
where
\begin{equation}
\label{eq:matr_R}
\vc{R}(r)=
\begin{pmatrix}
j_j(\rho_1)&h_j^{(1)}(\rho_1)&0&0\\
n_1\mu_1^{-1}Dj_j(\rho_1)&
n_1\mu_1^{-1}Dh_j^{(1)}(\rho_1)&0&0\\
0&0&
j_{\tilde{j}}(\rho_1)&h_{\tilde{j}}^{(1)}(\rho_1)\\
0&0&
-\mu_1 n_{1}^{-1}
Dj_{\tilde{j}}(\rho_1)&
-\mu_1 n_{1}^{-1}
Dh_{\tilde{j}}^{(1)}(\rho_1)\\
\end{pmatrix}\, .
\end{equation}
\end{widetext}
We now apply the boundary conditions at the dielectric discontinuities on the 
inside and outside  of the anisotropic layer. In the matrix notation
this yields:
\begin{widetext} 
\begin{subequations}
\label{eq:sys1}
\begin{align}
\vc{R}_2
\begin{pmatrix}
\alpha_{jm}\\
\beta_{jm}\\
\tilde{\alpha}_{jm}\\
\tilde{\beta}_{jm}
\end{pmatrix}
&=\alpha_{jm}^{(c)}
\begin{pmatrix}
[j_j(\rho_2)]_2\\
n_2\mu_2^{-1}[j_j(\rho_2)]'_2\\
0\\
0 \end{pmatrix}
+\tilde{\alpha}_{jm}^{(c)}
\begin{pmatrix}
0\\
0\\
\left[j_{j}(\rho_2)\right]_2\\
-n_2^{-1}\mu_2 [j_j(\rho_2)]'_2 
\end{pmatrix}
\label{eq:sys1.a}\\
\vc{R}_1
\begin{pmatrix}
\alpha_{jm}\\
\beta_{jm}\\
\tilde{\alpha}_{jm}\\
\tilde{\beta}_{jm}
\end{pmatrix}
& =\beta_{jm}^{(sca)}
\begin{pmatrix}
[h_j^{(1)}(\rho)]_1\\
n/\mu\, [h_j^{(1)}(\rho)]'_1\\
 0\\
0
\end{pmatrix}
+\tilde{\beta}_{jm}^{(sca)}
\begin{pmatrix}
0\\
0\\
\,[h_j^{(1)}(\rho)]_1\\
-\mu/n\,\left[h_j^{(1)}(\rho)\right]'_1
\end{pmatrix}+\notag\\
& +\alpha_{jm}^{(inc)}
\begin{pmatrix}
[j_j(\rho)]_1\\
n/\mu\, [j_j(\rho)]'_1\\
 0\\
0
\end{pmatrix}
+\tilde{\alpha}_{jm}^{(inc)}
\begin{pmatrix}
0\\
0\\
\,[j_j(\rho)]_1\\
-\mu/n\,\left[j_j(\rho)\right]'_1
\end{pmatrix}\, ,
\label{eq:sys1.b}
\end{align}
\end{subequations}
\end{widetext}
where  
$\vc{R}_i\equiv \vc{R}|_{r=R_i}$, 
$D f(x)|_{r=R_i}\equiv[f(x)]'_i$ and $f(x)|_{r=R_i}\equiv[f(x)]_i$.

The amplitude in the anisotropic layer,
$\{\alpha_{jm},\beta_{jm},\tilde{\alpha}_{jm},\tilde{\beta}_{jm}\}$,
can be related to the amplitudes inside the core,
$\{\alpha_{jm}^{(c)},\tilde{\alpha}_{jm}^{(c)}\}$, by 
multiplying both sides of Eq.~\eqref{eq:sys1.a} by
$\vc{R}_2^{-1}$.
Substituting this result into the left hand side of
Eq.~\eqref{eq:sys1.b} enables the amplitudes inside the anisotropic layer 
to be eliminated, resulting now in a system of four equations
for the six amplitudes
$\{\alpha_{jm}^{(c)}$, $\tilde{\alpha}_{jm}^{(c)}$,
$\beta_{jm}^{(sca)}$, $\tilde{\beta}_{jm}^{(sca)}$,
$\alpha_{jm}^{(inc)}$, $\tilde{\alpha}_{jm}^{(inc)}\}$:
\begin{widetext}
\begin{align}
&
\begin{pmatrix}
c_{11}& c_{12}& -[h_j^{(1)}(\rho)]_1 & 0\\
c_{21}& c_{22}& -n/\mu\, [h_j^{(1)}(\rho)]'_1 & 0\\
c_{31}& c_{32}& 0 &-[h_j^{(1)}(\rho)]_1\\
c_{41}& c_{42}& 0 &\mu/n\,[h_j^{(1)}(\rho)]'_1
\end{pmatrix}
\begin{pmatrix}
\alpha_{jm}^{(c)}\\
\tilde{\alpha}_{jm}^{(c)}\\
\beta_{jm}^{(sca)}\\
\tilde{\beta}_{jm}^{(sca)}\\
\end{pmatrix}=\notag\\
&=\alpha_{jm}^{(inc)}
\begin{pmatrix}
\,[j_j(\rho)]_1\\
n/\mu\, [j_j(\rho)]'_1\\
 0\\
0
\end{pmatrix}
+\tilde{\alpha}_{jm}^{(inc)}
\begin{pmatrix}
0\\
0\\
\,[j_j(\rho)]_1\\
-\mu/n\,[j_j(\rho)]'_1
\end{pmatrix}\, ,
\label{eq:sys2}
\end{align}
where
\begin{equation}
\label{eq:c}
\vc{C}\equiv
\begin{pmatrix}
c_{11}& c_{12}\\
c_{21}& c_{22}\\
c_{31}& c_{32}\\
c_{41}& c_{42}
\end{pmatrix}=
\vc{R}_1\vc{R}_2^{-1}
\begin{pmatrix}  
[j_j(\rho_2)]_2 & 0\\
n_2\mu_2^{-1}[j_j(\rho_2)]'_2 & 0\\
 0 & [j_j(\rho_2)]_2\\
0 & -n_2^{-1}\mu_2 [j_j(\rho_2)]'_2
\end{pmatrix}\, .
\end{equation}
\end{widetext}
We now eliminate $\{\alpha_{jm}^{(c)}$, $\tilde{\alpha}_{jm}^{(c)}\}$ in
the system~\eqref{eq:sys2}. 
This yields the standard general form for the relation between the
amplitudes of the  
incoming and outgoing waves ~\eqref{eq:matr}, with the 
following explicit expressions for the elements of
the $T$--matrix:
\begin{widetext}
\begin{equation}
  \label{eq:t_matr_rad}
T_j^{11}=\frac{c_2\,[j_j(\rho)]_1-
{n}/{\mu}\,c_1\,[j_j(\rho)]'_1}
{{n}/{\mu}\,c_1\,[h_j^{(1)}(\rho)]'_1-
c_2\,[h_j^{(1)}(\rho)]_1}\, ,\quad
T_j^{22}=-\frac{\tilde{c}_2\,[j_j(\rho)]_1+
{\mu}/{n}\,\tilde{c}_1\,[j_j(\rho)]'_1}
{{\mu}/{n}\,\tilde{c}_1\,[h_j^{(1)}(\rho)]'_1+
\tilde{c}_2\,[h_j^{(1)}(\rho)]_1}\, ,
\end{equation}
\end{widetext}
where $c_i\equiv c_{i\,1}$ and $\tilde{c}_i\equiv c_{i+2,\,2}$.

We note that in this result the off-diagonal elements of the
$T$--matrix vanish: $T_j^{12}=T_j^{21}=0$. This result is quite
general for spherically symmetric scatterers, and physically means
that there is no coupling between transverse electric and transverse
magnetic waves. Equivalently there is no depolarisation scattering in
this scatterer.

Mathematically the result follows because
the matrix~\eqref{eq:matr_R} is block diagonal. Thus 
from Eq.~\eqref{eq:c} we have
$c_{31}=c_{41}=c_{12}=c_{22}=0$, from which the result follows.
The \T--matrix is diagonal over the angular momenta $j$ 
and the azimuthal numbers $m$:
$T_{jj',mm'}^{nn'}=\delta_{jj'}\delta_{mm'}
\delta_{nn'}T_{j}^{nn}$.

\section{Scattering from uniformly anisotropic layer}
\label{sec:scatt-uni-anis}

By contrast  with the case of the radially anisotropic scatterer
considered in the previous section, the electromagnetic field
inside the uniformly anisotropic annular layer can  easily be
described in terms of plane waves. Unfortunately this does not render
the scattering problem soluble. The difficulty lies in satisfying
the boundary conditions for a spherical scatterer 
using the plane wave solutions of the Maxwell equations.

The starting point of the \T-matrix approach to this case involves
finding a  representation of the electromagnetic wave 
analogous to the generalised \T-matrix ansatz~\eqref{eq:anis}.  
However, for symmetry reasons, the structure of the modes
representing the fields in the uniformly anisotropic medium
differs from that in the spherically symmetric isotropic
(or a radially anisotropic) geometry.
In particular, the lack of spherical symmetry implies that
the angular momentum number $j$ is no longer a good quantum
number. However, the cylindrical 
symmetry still guarantees conservation of the azimuthal number $m$.

The procedure is as follows. In Sec.~\ref{subsec:ang-moment-ani}
we provide methods of defining  modes in a uniformly anisotropic
material. These modes are: (a)~solutions of the Maxwell's equations and
(b)~deformations of the isotropic spherical harmonics. 
The latter condition means that  the isotropic modes  Eqs.~\eqref{eq:iso}
are recovered in the weak anisotropy limit, $u\to 0$.
Then in Sec.~\ref{subsec:t-matrix-uni}, 
we shall use these ``quasi-spherical'' wave functions
to derive equations which enable  the elements of \T-matrix to be computed
for a uniformly anisotropic annular layer.

\subsection{Angular Momentum Representation 
in the Anisotropic Layer}
\label{subsec:ang-moment-ani}

We have seen above that solutions to Maxwell's equations can be
written in two ways.  Either they can be expressed as plane waves, or,
using a separation of variables approach, they can be written as
expansions over spherical harmonics. Deriving an expression for the
\T--matrix will require us to make a connection between these two
alternative expansions.  In this subsection we carry out this task.

To do this we begin with an isotropic medium. In this special case
both the spherical harmonics and the plane wave solutions are known.
The result is a relation between the plane wave packets and the
spherical harmonics. The procedure will then be generalised to cover
the case of a uniformly anisotropic medium, so as to derive a set of
"quasi-spherical" normal modes.

\subsubsection{Spherical modes and plane waves in isotropic media}
\label{subsubsec:iso-medium}

We start with the Maxwell equations~\eqref{eq:maxwell} for an
isotropic medium.  
Wave-like solutions to these equations written  in terms of 
spherical coordinate basis functions are given by
Eqs.~\eqref{eq:sep} and~\eqref{eq:iso}.  
An electromagnetic wave can alternatively be written as a  superposition of
plane waves:
\begin{widetext}
\begin{subequations}
  \label{eq:iso_pl}
\begin{align}
  \vc{E}=\langle\exp(i\rho\, \uvc{k}\cdot\uvc{r})\,
\bigl[E_x(\uvc{k})\,\vc{e}_x(\uvc{k})+
E_y(\uvc{k})\,\vc{e}_y(\uvc{k})
\bigr]\rangle_{\uvc{k}}\, ,
  \label{eq:iso_pl_e}\\
  \vc{H}=\frac{n}{\mu}\,\langle\exp(i\rho\, \uvc{k}\cdot\uvc{r})\,
\bigl[E_x(\uvc{k})\,\vc{e}_y(\uvc{k})-
E_y(\uvc{k})\,\vc{e}_x(\uvc{k})
\bigr]\rangle_{\uvc{k}}\, ,
  \label{eq:iso_pl_h}
\end{align}
\end{subequations}
\end{widetext}
where 
$\displaystyle
\langle\,f\,\rangle_{\uvc{k}}\equiv\int_0^{2\pi}\dd\phi_k
\int_0^{\pi}\sin\theta_k\dd\theta_k\,f$.

In Appendix B we derive expressions connecting the isotropic
spherical modes~\eqref{eq:iso1} 
and the vector plane  waves occurring in the  superposition~\eqref{eq:iso_pl}:
\begin{widetext}
\begin{subequations}
\label{eq:wave_fun_iso}
\begin{align}
& 
\vc{M}^{(m)}_{jm}(\rho,\uvc{r})=\notag\\
&=i^{-j}(4\pi)^{-2}\sqrt{\pi(2j+1)}\,
\langle\,
\exp(i\rho\, \uvc{k}\cdot\uvc{r})
\left[\, 
D_{jm}^{(y)\,*}(\uvc{k})\,
\vc{e}_x(\uvc{k})
-i D_{jm}^{(x)\,*}(\uvc{k})
\,\vc{e}_y(\uvc{k})
\,\right]
\,\rangle_{\uvc{k}}\, ,
\label{eq:wave_fun_m_m}\\
& 
\vc{M}^{(e)}_{jm}(\rho,\uvc{r})=\notag\\
&=i^{-j}(4\pi)^{-2}\sqrt{\pi(2j+1)}\,
\langle\,
\exp(i\rho\, \uvc{k}\cdot\uvc{r})
\left[\, 
i D_{jm}^{(x)\,*}(\uvc{k})\,
\vc{e}_x(\uvc{k})
+ D_{jm}^{(y)\,*}(\uvc{k})
\,\vc{e}_y(\uvc{k})
\,\right]
\,\rangle_{\uvc{k}}\, ,
\label{eq:wave_fun_m_e}
\end{align}
\end{subequations}
where
\begin{subequations}
  \label{eq:funcs}
\begin{gather}
  D^{(x)}_{jm}(\uvc{k})\equiv\exp(-im\,\phi_k)\,d^{(x)}_{jm}(\theta_k)=
D^j_{m,-1}(\uvc{k})-D^j_{m,\,1}(\uvc{k})\, ,
\label{eq:d_x}\\
  D^{(y)}_{jm}(\uvc{k})\equiv\exp(-im\,\phi_k)\,d^{(y)}_{jm}(\theta_k)=
D^j_{m,-1}(\uvc{k})+D^j_{m,\,1}(\uvc{k})\, ,\quad
D^{(z)}_{jm}(\uvc{k})\equiv D^j_{m,\,0}(\uvc{k})\, .
\label{eq:d_y_z}
\end{gather}
\end{subequations}
\end{widetext}

These relations can be explicitly verified  by
substituting the expansions~\eqref{eq:plane_iso} into
the right hand sides of Eqs.~\eqref{eq:wave_fun_iso}.
The linear combinations of the modes $\vc{M}^{(m)}_{jm}(\rho,\uvc{r})$
and $\vc{M}^{(e)}_{jm}(\rho,\uvc{r})$ which enter
the  electromagnetic field harmonics~\eqref{eq:sep}
can now be expressed as a superposition of plane waves:
\begin{widetext}
\begin{subequations}
\label{eq:sep_plw}
\begin{align}
& 
\alpha_{jm}\vc{M}^{(m)}_{jm}(\rho,\uvc{r})
-\frac{\mu}{n}
\tilde\alpha_{jm}\vc{M}^{(e)}_{jm}(\rho,\uvc{r})
=\langle\exp(i\rho\, \uvc{k}\cdot\uvc{r})\,
\bigl[E_{jm}^{(x)}(\uvc{k})\,\vc{e}_x(\uvc{k})+
E_{jm}^{(y)}(\uvc{k})\,\vc{e}_y(\uvc{k})
\bigr]\rangle_{\uvc{k}}\, ,
\label{eq:sep_plw1}\\
& 
\tilde\alpha_{jm}\vc{M}^{(m)}_{jm}(\rho,\uvc{r})+
\frac{n}{\mu}
\alpha_{jm}\vc{M}^{(e)}_{jm}(\rho,\uvc{r})
=\frac{n}{\mu}\,\langle\exp(i\rho\, \uvc{k}\cdot\uvc{r})\,
\bigl[E_{jm}^{(x)}(\uvc{k})\,\vc{e}_y(\uvc{k})-
E_{jm}^{(y)}(\uvc{k})\,\vc{e}_x(\uvc{k})
\bigr]\rangle_{\uvc{k}}\, ,
\label{eq:sep_plw2}
\end{align} 
\end{subequations}
where
\begin{subequations}
  \label{eq:basis}
\begin{align}
E^{(x)}_{jm}(\uvc{k})=\frac{i^{-j}}{(4\pi)^2}\,[\pi(2j+1)]^{1/2}
\left\{
\alpha_{jm}\,D^{(y)\,*}_{jm}(\uvc{k}) -i\,
\frac{\mu}{n}\,\tilde\alpha_{jm}\,
D^{(x)\,*}_{jm}(\uvc{k})
\right\}\, ,
  \label{eq:basis1}\\
E^{(y)}_{jm}(\uvc{k})=-\frac{i^{-j}}{(4\pi)^2}\,[\pi(2j+1)]^{1/2}
\left\{
i\,\alpha_{jm}\,D^{(x)\,*}_{jm}(\uvc{k}) +
\frac{\mu}{n}\,\tilde\alpha_{jm}\,
D^{(y)\,*}_{jm}(\uvc{k})
\right\}\, . 
  \label{eq:basis2}
\end{align}
\end{subequations}
\end{widetext}
We now  sum  Eqs.~\eqref{eq:sep_plw} over $j$ and $m$. 
This enables the amplitudes $E_{x}(\uvc{k})$ and $E_{y}(\uvc{k})$  in
Eqs.~\eqref{eq:iso_pl} 
to be expressed in terms of Wigner $D$-functions: 
\begin{widetext}
\begin{equation}
 E_{x}(\uvc{k})=\sum_{jm} E^{(x)}_{jm}(\uvc{k})\, ,\quad
 E_{y}(\uvc{k})=\sum_{jm} E^{(y)}_{jm}(\uvc{k})\, .
  \label{eq:basis0}
\end{equation}
\end{widetext}
This procedure has  started from plane waves~\eqref{eq:iso_pl} 
and spherical harmonics~\eqref{eq:sep}, and finished with
Eqs.~\eqref{eq:basis}-~\eqref{eq:basis0}. This equation defines a basis set
in the space of  the angular dependent amplitudes.

In fact we shall need to carry out the inverse process. 
The inverse process uses the expansions~\eqref{eq:basis0}
to derive the expressions for the spherical modes
$\vc{M}^{(\alpha)}_{jm}$ and
$\tilde{\vc{M}}^{(\alpha)}_{jm}$
from superpositions of plane waves~\eqref{eq:iso_pl}.  
The procedure works as follows:
\renewcommand{\theenumi}{\alph{enumi}}
\renewcommand{\labelenumi}{(\theenumi)}
\begin{enumerate}
\item
We substitute the expansions of the amplitudes $E_{x}(\uvc{k})$ and
$E_{y}(\uvc{k})$ from Eqs.~\eqref{eq:basis0} 
into the superpositions~\eqref{eq:iso_pl}.

\item
From the expressions for the electric (magnetic) fields we obtain 
the spherical modes  as coefficient
functions proportional to $\alpha_{jm}$ 
and $-\mu/n\,\tilde{\alpha}_{jm}$
($\tilde\alpha_{jm}$ and $n/\mu\,{\alpha}_{jm}$).
 
\item
In order to deduce explicit analytical expressions for the modes,
we expand the plane waves over vector spherical functions
by using Eqs.~\eqref{eq:plane_iso}. We then
integrate the products of Wigner $D$--functions over the angles
$\phi_k$ and $\theta_k$ by using 
the orthogonality condition~\eqref{eq:A.9}.

\item 
Finally, the modes $\tilde{\vc{M}}^{(\alpha)}_{jm}$ are derived
from the expressions for $\vc{M}^{(\alpha)}_{jm}$ by changing the
Bessel functions, $j_j(\rho)$, to the Hankel functions,
$h^{(1)}_j(\rho)$.
\end{enumerate}
Note that, if a linear combination of Bessel functions,
$j_j(\rho)$, represents a solution of linear homogeneous differential
equations (Maxwell equations in our case), then the corresponding
linear combination of Hankel functions generates another solution.
This remark justifies the last step in the  procedure described above.

The crucial point is that this inverse procedure can be generalised 
to a uniformly anisotropic medium.
Thus it can be applied to superpositions
of plane waves representing solutions of the Maxwell's equations in
the uniformly anisotropic layer,  yielding expressions for the modes
used in the generalised \T-matrix ansatz~\eqref{eq:anis}. In the 
next subsection we perform this generalisation.

\subsubsection{Wave functions in an anisotropic medium}
\label{subsubsec:wav-fun-uni}

We start with the expansion~\eqref{eq:basis0}, and use it to derive
formulae for 
generalised spherical harmonics in the anisotropic medium.
The starting point is the well known result 
for plane waves~\cite{Born,Landau:el,Lax:1971,Stark:1997}:
\begin{widetext}
\begin{subequations}
  \label{eq:anis_pl}
\begin{align}
 \vc{E}=\langle\exp(i\rho_e \uvc{k}\cdot\uvc{r})
E_x(\uvc{k})\bigl[
\vc{e}_x(\uvc{k})+\frac{u}{1+u}\,\sin\theta_k
\,\uvc{z}\bigr]+
\exp(i\rho\,\uvc{k}\cdot\uvc{r})
E_y(\uvc{k})\,\vc{e}_y(\uvc{k})
\rangle_{\uvc{k}}\, ,
  \label{eq:anis_pl1}\\
 \vc{H}=\frac{n}{\mu}\,\langle\exp(i\rho_e\uvc{k}\cdot\uvc{r})\,
n_e^{-1}\,E_x(\uvc{k})\vc{e}_y(\uvc{k})-
\exp(i\rho\,\uvc{k}\cdot\uvc{r})
E_y(\uvc{k})\vc{e}_x(\uvc{k})
\rangle_{\uvc{k}}\, ,
  \label{eq:anis_pl2}
\end{align}
\end{subequations}
\end{widetext}
where
$n_e^2\equiv n_e^2(\theta_k)=\dfrac{1+u}{1+u\cos^2\theta_k}\,$
and $\rho_e\equiv n_e(\theta_k)\,\rho$.

We now apply the procedure described at the
end of the last section to the plane wave packets~\eqref{eq:anis_pl}.
This gives a representation of the electromagnetic field in the form
of the generalised \T-matrix ansatz~\eqref{eq:anis}. Now, however,
the modes no longer take the form~\eqref{eq:rad_bas}, but rather the
modified form: 
\begin{widetext}
\begin{subequations}
\label{eq:wave_fun}
\begin{align}
& \vc{P}^{(m)}_{jm}(\rho,\uvc{r})=
i^{-j}(4\pi)^{-2}\sqrt{\pi(2j+1)}\,
\langle\, D_{jm}^{(y)\,*}(\uvc{k})\,
\exp(i\rho_e \uvc{k}\cdot\uvc{r})
\bigl[
\vc{e}_x(\uvc{k})+\frac{u}{1+u}\,\sin\theta_k
\,\uvc{z}\bigr] -\notag\\ 
&\phantom{%
\vc{P}^{(m)}_{jm}(\rho,\uvc{r})=
\sqrt{\pi(2j+1)}
\langle\, D_{jm}^{(y)\,*}(\uvc{k})\,
\exp(i\rho_e \uvc{k}\cdot\uvc{r})
}
-i D_{jm}^{(x)\,*}(\uvc{k})
\exp(i\rho\,\uvc{k}\cdot\uvc{r})
\,\vc{e}_y(\uvc{k})\,
\rangle_{\uvc{k}}\, ,
\label{eq:wave_fun_p_m}\\
& \vc{P}^{(e)}_{jm}(\rho,\uvc{r})=
i^{-j}(4\pi)^{-2}\sqrt{\pi(2j+1)}\,
\langle\, i D_{jm}^{(x)\,*}(\uvc{k})\,
\exp(i\rho_e \uvc{k}\cdot\uvc{r})
\bigl[
\vc{e}_x(\uvc{k})+\frac{u}{1+u}\,\sin\theta_k
\,\uvc{z}\bigr] +\notag\\ 
&\phantom{%
\vc{P}^{(m)}_{jm}(\rho,\uvc{r})=
\sqrt{\pi(2j+1)}
\langle\, D_{jm}^{(y)\,*}(\uvc{k})\,
\exp(i\rho_e \uvc{k}\cdot\uvc{r})
}
+ D_{jm}^{(y)\,*}(\uvc{k})
\exp(i\rho\,\uvc{k}\cdot\uvc{r})
\,\vc{e}_y(\uvc{k})\,
\rangle_{\uvc{k}}\, ,
\label{eq:wave_fun_p_e}\\
& \vc{Q}^{(m)}_{jm}(\rho,\uvc{r})=
i^{-j}(4\pi)^{-2}\sqrt{\pi(2j+1)}\,
\langle\, -i D_{jm}^{(x)\,*}(\uvc{k})\,
\exp(i\rho_e \uvc{k}\cdot\uvc{r})
\,n_e^{-1}\,
\vc{e}_y(\uvc{k})+\notag\\ 
&\phantom{%
\vc{P}^{(m)}_{jm}(\rho,\uvc{r})=
\sqrt{\pi(2j+1)}
\langle\, D_{jm}^{(y)\,*}(\uvc{k})\,
\exp(i\rho_e \uvc{k}\cdot\uvc{r})
}
+ D_{jm}^{(y)\,*}(\uvc{k})
\exp(i\rho\,\uvc{k}\cdot\uvc{r})
\,\vc{e}_x(\uvc{k})\,
\rangle_{\uvc{k}}\, ,
\label{eq:wave_fun_q_m}\\
& \vc{Q}^{(e)}_{jm}(\rho,\uvc{r})=
i^{-j}(4\pi)^{-2}\sqrt{\pi(2j+1)}\,
\langle\, D_{jm}^{(y)\,*}(\uvc{k})\,
\exp(i\rho_e \uvc{k}\cdot\uvc{r})
\,n_e^{-1}\,
\vc{e}_y(\uvc{k})+\notag\\ 
&\phantom{%
\vc{P}^{(m)}_{jm}(\rho,\uvc{r})=
\sqrt{\pi(2j+1)}
\langle\, D_{jm}^{(x)\,*}(\uvc{k})\,
\exp(i\rho_e \uvc{k}\cdot\uvc{r})
}
+ i D_{jm}^{(y)\,*}(\uvc{k})
\exp(i\rho\,\uvc{k}\cdot\uvc{r})
\,\vc{e}_x(\uvc{k})\,
\rangle_{\uvc{k}}\, ,
\label{eq:wave_fun_q_e}
\end{align} 
\end{subequations}
\end{widetext}

Typically, solving a scattering problem for a spherical particle
requires  modes to be expressed in terms of vector spherical harmonics.
It thus turns out to be useful to write the wave
functions~\eqref{eq:wave_fun} as expansions
over vector spherical harmonics:

\begin{widetext}
\begin{subequations}
  \label{eq:uni}
\begin{align}
&  
\vc{P}^{(\alpha)}_{jm}=\sum_{\beta}\sum_{j'\ge\, |m|}
p_{j'j;\,m}^{\,(\beta,\,\alpha)}(\rho)
\vc{Y}_{j'm}^{\,(\beta)}(\uvc{r})\, ,\quad
\tilde{\vc{P}}^{(\alpha)}_{jm}=\sum_{\beta}\sum_{j'\ge\, |m|}
\tilde{p}_{j'j;\,m}^{\,(\beta,\,\alpha)}(\rho)
\vc{Y}_{j'm}^{\,(\beta)}(\uvc{r})\, ,
\label{eq:uni1}\\
&
\vc{Q}^{(\alpha)}_{jm}=\sum_{\beta}\sum_{j'\ge\, |m|}
q_{j'j;\,m}^{\,(\beta,\,\alpha)}(\rho)
\vc{Y}_{j'm}^{\,(\beta)}(\uvc{r})\, ,\quad
\tilde{\vc{Q}}^{(\alpha)}_{jm}=\sum_{\beta}\sum_{j'\ge\, |m|}
\tilde{q}_{j'j;\,m}^{\,(\beta,\,\alpha)}(\rho)
\vc{Y}_{j'm}^{\,(\beta)}(\uvc{r})\, ,
\label{eq:uni2}
\end{align}
\end{subequations}
where $\alpha\in\{m,e\}$, $\beta\in\{m,e,o\}$ and
\begin{align}
&p_{j'j;\,m}^{\,(\beta,\,\alpha)}(\rho)=
  \langle \vc{Y}_{j'm}^{\,(\beta)\,*}(\uvc{r})\cdot
\vc{P}_{jm}^{\,(\alpha)}(\rho,\uvc{r})\rangle_{\uvc{r}}
\, ,\quad
\tilde{p}_{j'j;\,m}^{\,(\beta,\,\alpha)}(\rho)=
  \langle \vc{Y}_{j'm}^{\,(\beta)\,*}(\uvc{r})\cdot
\tilde{\vc{P}}_{jm}^{\,(\alpha)}(\rho,\uvc{r})\rangle_{\uvc{r}}\, ,
\notag\\
&q_{j'j;\,m}^{\,(\beta,\,\alpha)}(\rho)=
  \langle \vc{Y}_{j'm}^{\,(\beta)\,*}(\uvc{r})\cdot
\vc{Q}_{jm}^{\,(\alpha)}(\rho,\uvc{r})\rangle_{\uvc{r}}
\, ,\quad
\tilde{q}_{j'j;\,m}^{\,(\beta,\,\alpha)}(\rho)=
  \langle \vc{Y}_{j'm}^{\,(\beta)\,*}(\uvc{r})\cdot
\tilde{\vc{Q}}_{jm}^{\,(\alpha)}(\rho,\uvc{r})\rangle_{\uvc{r}}\, .
\label{eq:matr-R-uni-elem}
\end{align}
\end{widetext}

Explicit formulae for the coefficient functions entering
Eqs..(\ref{eq:uni2}) are given in Appendix~\ref{sec:coef-func} together
with some related comments.  Evaluating these coefficients involves
computing some products of Bessel spherical function and Wigner
$D$--functions and integrating these expressions over $\theta_k$.
Their numerical evaluation is relatively easy.

The coefficients with $j\ne j'$ describe angular momentum mixing. It
can be shown that these terms go to zero in the absence of anisotropy,
when $u=0$ and $n_e=1$.

The modes introduced in this subsection
can be used to derive expansions
for incident plane wave in the anisotropic medium.
In an isotropic material
the corresponding expansions are given by 
Eqs.~\eqref{eq:inc} and~\eqref{eq:coef_inc}.
In a uniformly anisotropic medium these expansions
take an exactly analogous form: 
\begin{widetext}
\begin{gather}
\vc{E}_{jm}^{(inc)}=
\alpha_{jm}^{(inc)}\,\vc{P}^{(m)}_{jm}(\rho,\uvc{r})
-\frac{\mu}{n}\,
\tilde\alpha_{jm}^{(inc)}\,\vc{P}^{(e)}_{jm}(\rho,\uvc{r})\notag\, ,\\
\vc{H}_{jm}^{(inc)}=
\tilde\alpha_{jm}^{(inc)}\,\vc{Q}^{(m)}_{jm}(\rho,\uvc{r})
+\frac{n}{\mu}\,
\alpha_{jm}^{(inc)}\,\vc{Q}^{(e)}_{jm}(\rho,\uvc{r})\, .
\label{eq:inc-ani}
\end{gather}
\end{widetext} 
These equations can be compared  to Eqs.~\eqref{eq:inc}.
In the anisotropic case  the spherical harmonics~\eqref{eq:iso} are replaced by
by the ``quasi-spherical'' modes~\eqref{eq:wave_fun}. 

In order to prove this result we insert the coefficients from
Eqs.~\eqref{eq:coef_inc} into
Eqs.~\eqref{eq:basis0}-\eqref{eq:basis2}.  Then, after performing some
rather straightforward algebraic manipulations, we obtain
\begin{widetext}
\begin{subequations}
\label{eq:compl}
\begin{align}
&E_{x}^{(inc)}(\uvc{k}_{inc},\uvc{k})=-2^{-1/2} 
\sum_{\nu=\pm 1} \sum_{jm}
(2j+1)/(4\pi)\,
\nu\,E_{\nu}^{(inc)}
D_{m\,\nu}^j(\uvc{k}_{inc})D_{m\,\nu}^{j\,*}(\uvc{k})\, ,
\label{eq:compl_x}\\
&E_{y}^{(inc)}(\uvc{k}_{inc},\uvc{k})=-i\, 2^{-1/2} 
\sum_{\nu=\pm 1} \sum_{jm}
(2j+1)/(4\pi)\,
E_{\nu}^{(inc)}
D_{m\,\nu}^j(\uvc{k}_{inc})D_{m\,\nu}^{j\,*}(\uvc{k})\, .
\label{eq:compl_y}
\end{align}
\end{subequations}
\end{widetext}

We finally observe that  the set
$\{\,D_{m\,\nu}^{j}(\uvc{k})\,\}_{jm}$ form an orthogonal 
(see Eq.~\eqref{eq:A.9} )and complete set of functions 
in the space of angular dependent functions.
The sums on the right hand sides of Eqs.~\eqref{eq:compl_x}
and~\eqref{eq:compl_y} can thus be written as angular $\delta$-functions:
\begin{widetext}
\begin{equation}
\sum_{jm}
(2j+1)/(4\pi)\,
D_{m\,\nu}^j(\uvc{k}_{inc})D_{m\,\nu}^{j\,*}(\uvc{k})=
\delta (\uvc{k}-\uvc{k}_{inc})\, .
\label{eq:delta-ang} 
\end{equation}
\end{widetext}

\subsection{\T-matrix: uniform anisotropy}
\label{subsec:t-matrix-uni}
 
In Sec.~\ref{subsec:t-matr-scatt} we solved for the \T-- matrix of a
radially anisotropic layer.  Computing the elements of \T-matrix
required the solution of a set of equations resulting from the
boundary conditions~\eqref{eq:cont}.  The only difference in other
cases is that that appropriate modes for the electromagnetic field
inside the anisotropic layer must be used. For uniform anisotropy
these modes are given by Eqs.~\eqref{eq:uni}. Mathematically,
Eqs.~\eqref{eq:matr_not} and~\eqref{eq:matr_R}, which describe the
fields in radially anisotropic layer, are replaced by the following
relations:
\begin{widetext}
\begin{equation}
  \label{eq:matr_not_uni}
\begin{pmatrix}
p_{jm}^{(m)}(r)\\
q_{jm}^{(e)}(r)\\
q_{jm}^{(m)}(r)\\
p_{jm}^{(e)}(r)
\end{pmatrix}
=\sum_{j'\ge |m|}\vc{R}^{jj';\,m}(r)\begin{pmatrix}
\alpha_{j'm}\\
\beta_{j'm}\\
\tilde{\alpha}_{j'm}\\
\tilde{\beta}_{j'm}
\end{pmatrix}\, ,
\end{equation}
where
\begin{align}
&\vc{R}^{jj';\,m}(r)=\notag\\
&=\begin{pmatrix}
p_{jj';\,m}^{\,(m,m)}(\rho_1)
&\tilde{p}_{jj';\,m}^{\,(m,m)}(\rho_1)
&-\mu_1 n_{1}^{-1}\,p_{jj';\,m}^{\,(m,e)}(\rho_1)
&-\mu_1 n_{1}^{-1}\,\tilde{p}_{jj';\,m}^{\,(m,e)}(\rho_1)
\\
n_1\mu_1^{-1}\,q_{jj';\,m}^{\,(e,e)}(\rho_1)
&n_1\mu_1^{-1}\,\tilde{q}_{jj';\,m}^{\,(e,e)}(\rho_1)
&q_{jj';\,m}^{\,(e,m)}(\rho_1)
&\tilde{q}_{jj';\,m}^{\,(e,m)}(\rho_1)
\\
n_1\mu_1^{-1}\,q_{jj';\,m}^{\,(m,e)}(\rho_1)
&n_1\mu_1^{-1}\,\tilde{q}_{jj';\,m}^{\,(m,e)}(\rho_1)
&q_{jj';\,m}^{\,(m,m)}(\rho_1)
&\tilde{q}_{jj';\,m}^{\,(m,m)}(\rho_1)
\\
p_{jj';\,m}^{\,(e,m)}(\rho_1)
&\tilde{p}_{jj';\,m}^{\,(e,m)}(\rho_1)
&-\mu_1 n_{1}^{-1}\,p_{jj';\,m}^{\,(e,e)}(\rho_1)
&-\mu_1 n_{1}^{-1}\,\tilde{p}_{jj';\,m}^{\,(e,e)}(\rho_1)
\\
\end{pmatrix}\, .
\label{eq:matr-R-uni}
\end{align}
\end{widetext}
These changes affect the left hand sides of
the system~\eqref{eq:sys1} which is modified in the following way
\begin{widetext}
\begin{subequations}
\label{eq:sys1-uni}
\begin{align}
\sum_{j'\,\ge\, |m|}
\vc{R}_2^{jj';\,m}
\begin{pmatrix}
\alpha_{j'm}\\
\beta_{j'm}\\
\tilde{\alpha}_{j'm}\\
\tilde{\beta}_{j'm}
\end{pmatrix}
&=\alpha_{jm}^{(c)}
\begin{pmatrix}
[j_j(\rho_2)]_2\\
n_2\mu_2^{-1}[j_j(\rho_2)]'_2\\
0\\
0 \end{pmatrix}
+\tilde{\alpha}_{jm}^{(c)}
\begin{pmatrix}
0\\
0\\
\left[j_{j}(\rho_2)\right]_2\\
-n_2^{-1}\mu_2 [j_j(\rho_2)]'_2 
\end{pmatrix}\, ,
\label{eq:sys1a-uni}\\
\sum_{j'\,\ge\, |m|}
\vc{R}_1^{jj';\,m}
\begin{pmatrix}
\alpha_{j'm}\\
\beta_{j'm}\\
\tilde{\alpha}_{j'm}\\
\tilde{\beta}_{j'm}
\end{pmatrix}
& =\beta_{jm}^{(sca)}
\begin{pmatrix}
[h_j^{(1)}(\rho)]_1\\
n/\mu\, [h_j^{(1)}(\rho)]'_1\\
 0\\
0
\end{pmatrix}
+\tilde{\beta}_{jm}^{(sca)}
\begin{pmatrix}
0\\
0\\
\, [h_j^{(1)}(\rho)]_1\\
-\mu/n\,[h_j^{(1)}(\rho)]'_1
\end{pmatrix}+\notag\\
& +\alpha_{jm}^{(inc)}
\begin{pmatrix}
[j_j(\rho)]_1\\
n/\mu\, [j_j(\rho)]'_1\\
 0\\
0
\end{pmatrix}
+\tilde{\alpha}_{jm}^{(inc)}
\begin{pmatrix}
0\\
0\\
\, [j_j(\rho)]_1\\
-\mu/n\,[j_j(\rho)]'_1
\end{pmatrix}\, .
\label{eq:sys1b-uni}
\end{align}
\end{subequations}
\end{widetext}

By analogy with the procedure adopted in
Sec.~\ref{subsec:t-matr-scatt}, this system of equations can now be
solved, in principle, to yield solutions for the quantities
$T^{\alpha,\beta}_{jj; m}$ which occur in
Eq.~\eqref{eq:matr_uni}. However, the crucial difficulty in this case
is that the algebraic structure of the equations is complicated by the
presence of angular momentum mixing.  Thus, by contrast with the
radially anisotropic layer considered in
Sec.~\ref{subsec:t-matr-scatt}, we are now unable to derive
expressions for the elements of the \T--matrix in closed form.  The
solution of this system of equations now requires numerical
analysis. In future publications in this series, we shall discuss this
problem in greater detail, and present some explicit results.

\begin{figure*}[!tbh]
\centering
\resizebox{150mm}{!}{\includegraphics{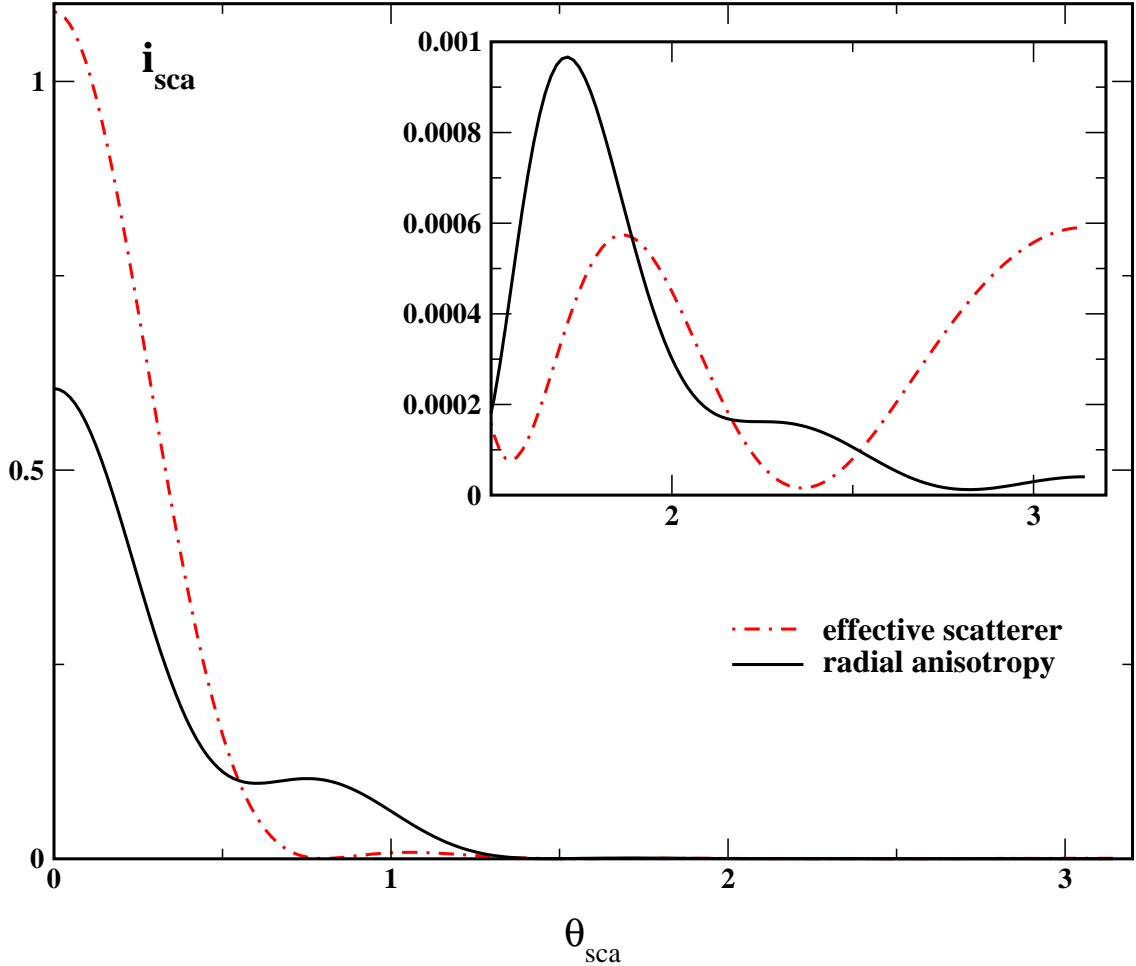}}
\caption{
Scattered intensity (see Eq.~\eqref{eq:ang-int})  
versus the scattering angle
at $kR_2=1.5$, $kd=4.0$  
and $u_1=0.25$ for (a)~effective isotropic scatterer with
$m_{\eff}\approx 1.05127$
($Q_{\eff}=0.14391$).  
and (b)~radially anisotropic layer, $\uvc{n}=\uvc{r}$
($Q_{Mie}=0.1439$). 
Insert at the upper right corner enlarges 
the backscattering tail.
}
\label{fig:eff-diff}
\end{figure*}

\begin{figure*}[!tbh]
\centering
\resizebox{150mm}{!}{\includegraphics{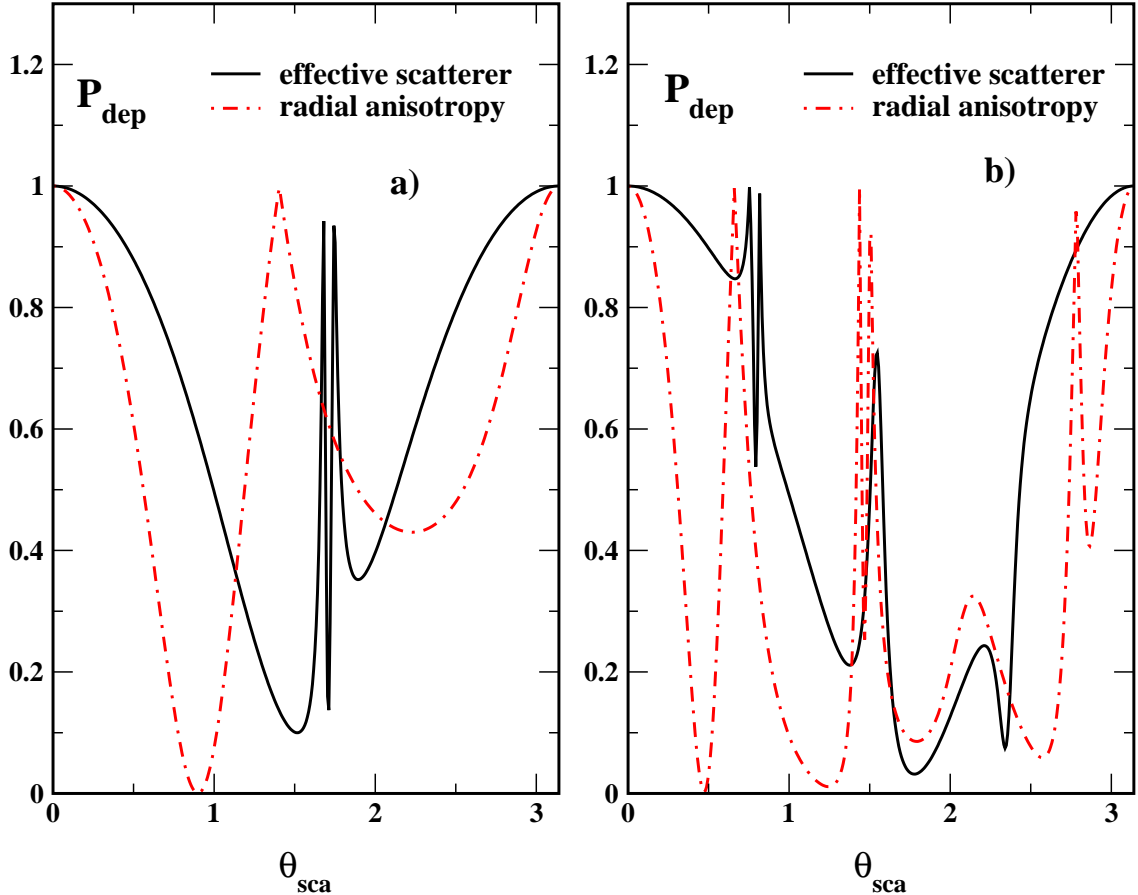}}
\caption{
Depolarisation factor (see Eq.~\eqref{eq:pol-fact})  
versus the scattering angle for 
both the radially anisotropic layer and its effective isotropic
scatterer at $kR_2=1.5$, $u_1=0.25$ and 
(a)~$kd=1.0$    
 ($m_{\eff}\approx 1.04512$, $Q_{\eff}=0.010724$,
$Q_{rad}=0.010729$),
(b)~$kd=4.0$  
($m_{\eff}\approx 1.05127$, $Q_{\eff}=0.14391$,
$Q_{rad}=0.1439$).
}
\label{fig:dep-eff-diff}
\end{figure*}

\section{Discussion and Conclusions}
\label{sec:concl}

In this paper we have developed a \T-matrix approach which can
describe light scattering by spherical scatterers containing optically
anisotropic material arranged in an annular layer.  We have confined
our discussion to the two cases, which we have called, using natural
language, radially and uniformly anisotropic systems. Just as in the
related case when the scatterer itself is anisotropic, but the
scattering material is optically isotropic, the presence of optical
anisotropy affects the algebraic structure which underlies the
\T-matrix theory.

From a mathematical point of view, the radial and uniform anisotropies present 
interesting but contrasting features. Here we draw the reader's attention to 
some of the most striking of these.  

The simplest case is the radially anisotropic layer. Here the
scattering material is locally optically anisotropic, but because of
the way that the anisotropy is arranged, the scatterer itself is
spherically symmetric and remains a globally optically isotropic
object.  The result is that there is no angular momentum mixing,
although we do have to introduce the new normal mode
structure~\eqref{eq:anis} within the anisotropic layer.

We shall discuss quantitative results in detail in a later paper in
this series.  However, it is our purpose here to show briefly that the
anisotropy effects can be important, even for this relatively simple
case.  As an example, we compare scattering by a radially anisotropic
layer and a radially isotropic layer of the same dimensions, chosen
in some some sense to be the best guess to an equivalent isotropic
scatterer.

Specifically, we consider scattering by a scatterer with a radially
anisotropic layer as discussed in section 2, with the
$\epsilon_{\perp}$ within the layer matching $\epsilon$ in the core
and outside the scatterer, and anisotropy coefficient $u$ inside the
layer. The equivalent effective scatterer has the same dimensions,
possesses an isotropic layer of refractive index $n_{\eff}$, so that
$\epsilon_{\eff}$ is the dielectric constant and $m_{\eff}=n_{\eff}/n$
is the optical contrast. The refractive index $n_{\eff}$ is chosen so
as to match scattering efficiencies of the anisotropic layer and of
the effective scatterer.

We compare the angular dependence of the scattering and the
depolarisation factors in the two cases.  The aggregate scattering is
the same, by definition. But disaggregated, we get different
contributions, both when looking at angles and when looking at
polarisation shifts.  In Fig.~\ref{fig:eff-diff} we compare the
angular dependence of the intensities. In the case we have considered,
the effective scatterer gives a much more forward scattering
signature, and the relatively tiny backward scattering contribution
has a very different angular structure.  Likewise, we see from
Fig.~\ref{fig:dep-eff-diff} that the depolarisation
factor~\eqref{eq:pol-fact} as a function of the scattering angle,
$\theta_{sca}$, is also very sensitive to the presence of anisotropy.

Given these relatively large effects for what might be thought of as minor 
anisotropy, there is every reason to suppose that the influence of 
a uniform anisotropic layer will be even more profound.
In this case the light-scattering problem is not exactly soluble.
The key point is that the \textit{exact solutions} for
uniformly anisotropic medium are known as plane waves,
whereas the spherical shape of the particle requires using some kind
of spherical modes.

In the case of the uniform anisotropy we have found it necessary to
examine the relation between spherical harmonic expansions and plane
wave solutions of Maxwell's equations.  We have found that, by choosing
the appropriate basis in $\uvc{k}$--space, we can define
'quasi-spherical' normal modes. These modes are \textit{exact
solutions} of Maxwell's equations and as such mix different angular
momentum. However, in the limit of zero anisotropy, these modes tend
to familiar spherical modes. More importantly, these quasi-spherical
modes turn out to be relatively easily accessible
computationally. Thus, there is every reason to suppose that the
strategy can be adopted in rather more complicated situations.

One such problem is the light scattering problem for a Faraday-active
sphere.  This problem has been treated using perturbation theory
in~\cite{Lacos:1998} to explain the origin of magneto-transverse light
diffusion known as the ``photonic Hall effect''~\cite{Tigg:1995,Tigg1:1996}.

We now try to place this problem in a more general physical
context. We were first motivated by the technological problem of
describing light transmission through media with liquid crystalline
inclusions, and the inverse problem in which the matrix is liquid
crystalline but the scatterers are isotropic. There is considerable
current interest in such materials for optical applications and
displays. The complete problem of light transmission through such
materials not only involves the single scattering processes discussed
in this paper, but also more general multiple scattering processes.

The \T--matrix formalism is a natural language within which to discuss
such problems.  beginning with single scattering theories of the type
discussed in this paper, one can in principle construct an effective
medium theory using, for example, the coherent potential approximation
(CPA) or coated CPA~\cite{Sheng:1992,Souk:1994a}.  These theories
determine effective optical characteristics of the medium from the
condition that the scattering cross section is minimal or equal to
zero on average.  Since this requires averaging over director
orientations, it is important to use basis functions with well defined
transformation properties under rotations.

This paper is the first in series of papers designed to improve
understanding of scattering by optically anisotropic bodies, both
singly and as components within complex anisotropic media.  In the
next papers in this series, we shall carry out detailed calculations
using the theory presented in this paper, and compare the results with
results derived using the simpler Rayleigh-Gans (RG) and Van de Hulst
or Anomalous Diffraction Approximations (ADA).

\begin{acknowledgments}
We acknowledge support from INTAS under grant 99--0312. AK thanks the
Faculty of Mathematical Studies in the University of Southampton for
its hospitality for a number of visits during 2000 and 2001.
\end{acknowledgments}

\appendix
\section{Vector spherical harmonics}
\label{sec:vect-spher-harm}

In this appendix we introduce notations and definitions used
throughout the paper.  In addition, we relate the vector spherical
harmonics and Wigner \textit{D}--functions.

Let us define the vectors
\begin{widetext} 
\begin{equation}
  \label{eq:A.1}
  \vc{e}_{\pm 1}(\uvc{r})=
\mp (\ubs{\vartheta}\pm i\ubs{\varphi})/\sqrt{2},\quad
  \vc{e}_0(\uvc{r})\equiv\uvc{r}\, ,
\end{equation}
\end{widetext}
where $\ubs{\varphi}=(-\sin\phi, \cos\phi, 0)$,
$\ubs{\vartheta}=
(\cos\theta\cos\phi, \cos\theta\sin\phi, -\sin\theta)$
are the unit vectors tangential to the sphere;
$\phi$ and $\theta$ are Euler angles of the unit vector $\uvc{r}$.
These vectors can be expressed in terms of the vectors of spherical basis, 
$\vc{e}_0\equiv\uvc{z}$,
$\vc{e}_{\pm 1}=\mp (\uvc{x}\pm i\uvc{y})/\sqrt{2}$, 
($\uvc{x},\,\uvc{y}$ and $\uvc{z}$ are the unit vectors directed along
the corresponding coordinate axes)
as follows:
\begin{widetext}
\begin{equation}
  \label{eq:A.2}
  \vc{e}_{\nu}(\uvc{r})=\sum_{\mu=-1}^{1}
  D_{\mu\nu}^{1}(\uvc{r})\,\vc{e}_{\mu},\quad \nu\in\{\pm 1,0\}\, ,
\end{equation}
where $D_{\mu\nu}^{\,j}(\uvc{r})\equiv D_{\mu\nu}^{\,j}(\phi,\theta)=
\exp(-i\mu\phi)\,d_{\mu\nu}^{\,j}(\theta)
$ is the Wigner $D$-function~\cite{Bie,Gel}.
The following properties of the vectors~\eqref{eq:A.2} are easy
to verify
\begin{equation}
  \label{eq:A.3}
(\vc{e}_{\nu}^{*}(\uvc{r})\cdot\vc{e}_{\mu}(\uvc{r}))=\delta_{\nu\mu},
\quad
\vc{e}_{\nu}(\uvc{r})\times\vc{e}_{-\nu}(\uvc{r})=
i\nu\,\vc{e}_{0}(\uvc{r})
\, .
\end{equation}
\end{widetext}

The vector spherical functions
${\vc{Y}}_{j m}^{\,(\alpha)}$ from Eqs.~\eqref{eq:spher}  
are expressed in terms of the vector spherical harmonics 
$\vc{Y}_{ljm}$~\cite{Bie} defined by
\begin{widetext}
\begin{equation}
  \label{eq:A.4}
  \vc{Y}_{l j m}=\sum_{\nu=-1}^{1} 
C_{m-\nu\, \nu\, m}^{l\,1\, j} 
\mathrm{Y}_{l\,m-\nu}\otimes\vc{e}_{\nu},\quad l=j+\delta,
\quad
\delta\in\{\pm 1, 0\}\, ,
\end{equation}
\end{widetext}
where $\mathrm{Y}_{lm}=\sqrt{(2l+1)/(4\pi)}
D_{m\,0}^{\,l\,*}$ is the spherical function~\cite{Abr} and
$C_{m-\nu\,\nu\, m}^{j+\delta\, 1\, j}$
denotes the Clebsch--Gordon (Wigner) coefficient,
\begin{widetext}
\begin{subequations}
  \label{eq:A.5}
\begin{align}  
& \vc{Y}_{jm}^{(e)}=s_j\vc{Y}_{j+1\,j\, m}+c_j\vc{Y}_{j-1\,j\,m},\quad
\vc{Y}_{jm}^{(m)}=\vc{Y}_{j\,j\,m}\\
& \vc{Y}_{jm}^{(o)}=-c_j\vc{Y}_{j+1\,j\,m}+s_j\vc{Y}_{j-1\,j\,m},\quad
s_j\equiv[j/(2j+1)]^{1/2},\, c_j\equiv[(j+1)/(2j+1)]^{1/2}\,.
\end{align}
\end{subequations}
\end{widetext}
Eqs.~\eqref{eq:A.2}, \eqref{eq:A.4} 
and the equality~\cite{Bie}
\begin{widetext}
\begin{gather}
  \label{eq:A.6}
    C_{k_1 k_2 k}^{j_1 j_2 j} D_{m k}^{\,j}=
\sum_{m_1 m_2}C_{m_1 m_2 m}^{j_1\, j_2\, j}
D_{m_1 k_1}^{\,j_1} D_{m_2 k_2}^{\,j_2}\, ,
\intertext{give}
\label{eq:A.7}
(\vc{Y}_{ljm}^{\,*}(\uvc{r})\cdot\vc{e}_{\nu}(\uvc{r}))=
\left[\frac{2j+1}{4\pi}\right]^{1/2}
C_{\nu}^{\,lj} D_{m\nu}^{\,j}(\uvc{r})\, ,
\end{gather}
where 
$C_{\nu}^{\,lj}\equiv [(2l+1)/(2j+1)]^{1/2}C_{0\,\nu\,\nu}^{\,l\,1\,j}\,$,
so that the non-vanishing values of $C_{\nu}^{\,lj}$ are
\begin{equation}
  \label{eq:A.8}
  \sqrt{2}\,C_{\nu}^{\,jj}=-\nu,\quad
  \sqrt{2}\,C_{\pm 1}^{\,j-1j}= -C_{0}^{\,j+1j}= c_j,\quad
  \sqrt{2}\,C_{\pm 1}^{\,j+1j}= C_{0}^{\,j-1j}= s_j\, .
\end{equation}
\end{widetext}

From Eqs.~\eqref{eq:A.5},~\eqref{eq:A.7} and~\eqref{eq:A.8}
we express  the vector spherical harmonics in terms of the
Wigner \textit{D}-functions as follows
\begin{widetext}
\begin{subequations}
  \label{eq:A.y2d}
\begin{align}
& \vc{Y}_{jm}^{(m)}(\uvc{r})=
[(2j+1)/8\pi]^{1/2}\left\{
D_{m,\,-1}^{j\,*}(\uvc{r})\,\vc{e}_{-1}(\uvc{r})-
D_{m,\,1}^{j\,*}(\uvc{r})\,\vc{e}_{+1}(\uvc{r})
\right\}\, ,
  \label{eq:A.y2d_m}\\
& \vc{Y}_{jm}^{(e)}(\uvc{r})=
[(2j+1)/8\pi]^{1/2}\left\{
D_{m,\,-1}^{j\,*}(\uvc{r})\,\vc{e}_{-1}(\uvc{r})+
D_{m,\,1}^{j\,*}(\uvc{r})\,\vc{e}_{+1}(\uvc{r})
\right\}\, ,
  \label{eq:A.y2d_e}\\
& \vc{Y}_{jm}^{(o)}(\uvc{r})=
[(2j+1)/4\pi]^{1/2}
D_{m,\, 0}^{j\,*}(\uvc{r})\,\vc{e}_{0}(\uvc{r})\, .
  \label{eq:A.y2d_o}
\end{align}
\end{subequations}
\end{widetext}

Note that the \textit{D}-functions meet the following
orthogonality relations~\cite{Bie,Gel}
\begin{widetext}
\begin{equation}
  \label{eq:A.9}
  \langle D_{m\nu}^{\,j\,*}(\uvc{r})
D_{m'\nu}^{\,j'}(\uvc{r})\rangle_{\uvc{r}}=
\frac{4\pi}{2j+1}\,\delta_{jj'}\,\delta_{mm'}\, ,
\end{equation}
where
$\displaystyle
\langle\,f\,\rangle_{\uvc{r}}\equiv\int_0^{2\pi}\dd\phi
\int_0^{\pi}\sin\theta\dd\theta\,f$. The orthogonality
condition~\eqref{eq:A.9} and Eqs.~\eqref{eq:A.y2d_m}-\eqref{eq:A.y2d_o}
show that a set of vector spherical harmonics is 
orthonormal:
\begin{equation}
  \langle \vc{Y}_{jm}^{(\alpha)\,*}(\uvc{r})\cdot
\vc{Y}_{j'm'}^{(\beta)}(\uvc{r})
\rangle_{\uvc{r}}= \delta_{jj'}\,\delta_{mm'}\, .
  \label{eq:A.10}
\end{equation}
\end{widetext}

\section{Rayleigh expansions for vector plane waves}
\label{sec:rayl-expans-vect}

In this Appendix we comment on the vector version of the
well known Rayleigh expansion (see, for example,~\cite{New}):
\begin{widetext}
\begin{equation}
  \label{eq:B.1}
  \exp(i\,\vc{k}\cdot\vc{r})=
4\pi\sum_{l=0}^{\infty}\sum_{m=-l}^{l} 
i^{\,l} j_l(\rho)\,Y_{lm}(\uvc{r})\,Y_{lm}^{*}(\uvc{k}),\quad
\rho\equiv kr\, .
\end{equation}
\end{widetext}

Let us consider plane wave with the wave vector
$\vc{k}\equiv k\uvc{k}$ and the polarisation vector $\vc{E}$
defined by its components, $E_{\nu}$, in the basis
$\vc{e}_{\nu}(\uvc{k})$ (see Eq.~\eqref{eq:A.2}):
\begin{widetext}
\begin{equation}
  \label{eq:vec_k}
   \vc{e}_{\nu}(\uvc{k})=\sum_{\mu=-1}^{1}
  D_{\mu\nu}^{1}(\uvc{k})\,\vc{e}_{\mu},\quad
\vc{e}_{\pm 1}(\uvc{k})=
\mp(\vc{e}_x(\uvc{k})\pm i\,\vc{e}_y(\uvc{k}))/\sqrt{2},\quad
\vc{E}=\sum_{\nu=-1}^{1}E_{\nu}\,\vc{e}_{\nu}(\uvc{k})\, ,
\end{equation}
$D_{\mu\nu}^{\,j}(\uvc{k})\equiv D_{\mu\nu}^{\,j}(\phi_k,\theta_k)=
\exp(-i\mu\phi_k)\,d_{\mu\nu}^{\,j}(\theta_k)$
is the Wigner $D$-function~\cite{Bie,Gel} and
$\phi_k$, $\theta_k$ are the azimuthal and polar angles of 
the unit vector $\uvc{k}$.
The vectors 
$\vc{e}_y(\uvc{k})\equiv\ubs{\varphi}_k=
(-\sin\phi_k, \cos\phi_k, 0)$ and
$\vc{e}_x(\uvc{k})\equiv\ubs{\vartheta}_k=
(\cos\theta_k\cos\phi_k, \cos\theta_k\sin\phi_k, -\sin\theta_k)$
are perpendicular to $\uvc{k}$.
\end{widetext}

From Eq.~\eqref{eq:B.1}, definition of the vector spherical functions
\eqref{eq:A.5} and the equality~\eqref{eq:A.6} it is not difficult
to derive the following relation
\begin{widetext}
\begin{equation}
  \label{eq:B.3}
  \vc{e}_{\nu}(\uvc{k})\exp(i\,\vc{k}\cdot\vc{r})=
\sum_{j,m}[2\pi(2j+1)]^{1/2}D_{m\nu}^{j}(\uvc{k})
\biggl[
\sum_l i^{\,l}\,j_l(\rho)\,C_{\nu}^{\,lj}\,\vc{Y}_{ljm}(\uvc{r})
\biggr]\, ,
\end{equation}
\end{widetext}
where $C_{\nu}^{\,lj}$ is defined in Eq.~\eqref{eq:A.8}.
The sum in square brackets on the right hand side of
Eq.~\eqref{eq:B.3} can be simplified by making use of 
Eq.~\eqref{eq:A.8} and the recursion relations~\cite{Abr}:
\begin{widetext}
\begin{equation}
  \label{eq:B.4}
  j_{j+1}+j_{j-1}=(2j+1)\rho^{-1}\,j_j,\qquad
  j_{j-1}-j_{j+1}=2\frac{\dd j_j}{\dd\rho}+\rho^{-1}\,j_j\, .
\end{equation}
\end{widetext}
The final result for transverse waves with
$\nu=\pm 1$ is
\begin{widetext}
\begin{align}
  \vc{e}_{\nu}\,(\uvc{k})&\exp(i\,\vc{k}\cdot\vc{r})=
\sum_{j,m}i^{\,j-1}\,[2\pi(2j+1)]^{1/2}\,D_{m\nu}^{j}(\uvc{k})\cdot\notag\\
&\cdot\left[
Dj_j(\rho)\vc{Y}_{jm}^{(e)}(\uvc{r})
-i\nu\, j_j(\rho)\vc{Y}_{jm}^{(m)}(\uvc{r})+
\sqrt{j(j+1)}\,\rho^{-1}\,j_j(\rho)\vc{Y}_{jm}^{(o)}(\uvc{r})
\right]\, .
\label{eq:B.5}
\end{align}
\end{widetext}

It can be written in the following form
\begin{widetext}
\begin{subequations}
  \label{eq:plane_iso}
\begin{align}  
  \vc{e}_x(\uvc{k})\exp(i\,\vc{k}\cdot\vc{r})=
\sum_{j,m}i^j\, [\pi(2j+1)]^{1/2}\left[
D^{(y)}_{jm}(\uvc{k})\,\vc{M}^{(m)}_{jm}(\rho,\uvc{r}) - i
D^{(x)}_{jm}(\uvc{k})\,\vc{M}^{(e)}_{jm}(\rho,\uvc{r})
\right]\, ,
  \label{eq:plane_iso1}\\
  \vc{e}_y(\uvc{k})\exp(i\,\vc{k}\cdot\vc{r})=
\sum_{j,m}i^j\, [\pi(2j+1)]^{1/2}\left[
i D^{(x)}_{jm}(\uvc{k})\,\vc{M}^{(m)}_{jm}(\rho,\uvc{r}) +
D^{(y)}_{jm}(\uvc{k})\,\vc{M}^{(e)}_{jm}(\rho,\uvc{r})
\right]\, ,
\label{eq:plane_iso2}
\end{align}
\end{subequations}
\end{widetext}
where the modes 
$\vc{M}_{jm}^{(m)}$, $\vc{M}_{jm}^{(e)}$ are defined
by Eq.~\eqref{eq:iso1} and 
the functions $D_{jm}^{(x)}$, $D_{jm}^{(y)}$
are expressed in terms of Wigner $D$-functions 
in Eqs.~\eqref{eq:d_x}-\eqref{eq:d_y_z}.

Let us consider the case when the polarisation vector
is directed along the $z$ axis. We derive the formulae for the 
following matrix elements:
$
\langle\,
\vc{Y}_{jm}^{(\alpha)\,*}\cdot
\uvc{z}\exp(i\,\vc{k}\cdot\vc{r})
\,\rangle_{\uvc{r}}
$.
In order to do it, notice that the definition~\eqref{eq:A.4}
immediately gives the following relation: 
\begin{equation}
\uvc{z}\,Y_{lm}(\uvc{r})=\sum_j
C^{\,l\,1\,j}_{m\,0\,m}
\vc{Y}_{ljm}\, .
\label{eq:B.6}
\end{equation}
Then we
substitute Eq.~\eqref{eq:B.6} into the expansion~\eqref{eq:B.1}
multiplied by $\uvc{z}$ and express the vector functions
$\vc{Y}_{ljm}$ in terms of the vector spherical harmonics
$\vc{Y}_{jm}^{(\alpha)}$. The result reads
\begin{widetext}
\begin{subequations}
\label{eq:B.7}
\begin{align}
&
\langle\,
\vc{Y}_{jm}^{(m)\,*}\cdot
\uvc{z}\,\exp(i\,\vc{k}\cdot\vc{r})
\,\rangle_{\uvc{r}}=
i^j [4\pi(2j+1)]^{1/2}
C^{\,j\,1\,j}_{m\,0\,m}\,
D^{\,j}_{m0}(\uvc{k})\,j_j(\rho)\, ,
\label{eq:B.z-pl-m}\\
&
\langle\,
\vc{Y}_{jm}^{(e)\,*}\cdot
\uvc{z}\,\exp(i\,\vc{k}\cdot\vc{r})
\,\rangle_{\uvc{r}}=
i^{j+1} [4\pi]^{1/2}
\Bigl[\,
(2j+3)^{1/2}\,s_j\,
C^{\,j+1\,1\,j}_{m\,0\,m}\,
D^{\,j+1}_{m0}(\uvc{k})\,j_{j+1}(\rho)-\notag\\
&
\phantom{
\langle\,
\vc{Y}_{jm}^{(e)\,*}\cdot
\uvc{z}\exp(i\,\vc{k}\cdot\vc{r})
\,\rangle_{\uvc{r}}=
}
-(2j-1)^{1/2}\,c_j\,
C^{\,j-1\,1\,j}_{m\,0\,m}\,
D^{\,j-1}_{m0}(\uvc{k})\,j_{j-1}(\rho)
\Bigr]\, ,
\label{eq:B.z-pl-e}\\
&
\langle\,
\vc{Y}_{jm}^{(o)\,*}\cdot
\uvc{z}\,\exp(i\,\vc{k}\cdot\vc{r})
\,\rangle_{\uvc{r}}=
i^{j+1} [4\pi]^{1/2}
\Bigl[\,
(2j+3)^{1/2}\,c_j\,
C^{\,j+1\,1\,j}_{m\,0\,m}\,
D^{\,j+1}_{m0}(\uvc{k})\,j_{j+1}(\rho)+\notag\\
&
\phantom{
\langle\,
\vc{Y}_{jm}^{(e)\,*}\cdot
\uvc{z}\exp(i\,\vc{k}\cdot\vc{r})
\,\rangle_{\uvc{r}}=
}
+(2j-1)^{1/2}\,s_j\,
C^{\,j-1\,1\,j}_{m\,0\,m}\,
D^{\,j-1}_{m0}(\uvc{k})\,j_{j-1}(\rho)
\Bigr]\, .
\label{eq:B.z-pl-o}
\end{align}
\end{subequations}
with the Wigner coefficients given by~\cite{Abr,Bie}
\begin{equation}
C^{\,j-1\,1\,j}_{m\,0\,m}=
\left[
\frac{j^2-m^2}{j(2j-1)}
\right]^{1/2}\, ,\,
C^{\,j\,1\,j}_{m\,0\,m}=
\frac{m}{\sqrt{j(j+1)}}\, ,\,
C^{\,j+1\,1\,j}_{m\,0\,m}=
- \left[
\frac{(j+1)^2-m^2}{(j+1)(2j+3)}
\right]^{1/2}\, .
\label{eq:B.8}
\end{equation}
\end{widetext}

In conclusion, note that in case, when a series on
the right hand side of Eq.~\eqref{eq:B.5} or Eqs.~\eqref{eq:B.7}
represent a solution of the Maxwell equations~\eqref{eq:maxwell}, 
we can obtain another
solution by replacing $j_j(\rho)$ with $h_j^{(1)}(\rho)$.

\section{Coefficient functions}
\label{sec:coef-func}

By definition, the coefficient functions that enter the
expansions~\eqref{eq:uni} are the matrix
elements~\eqref{eq:matr-R-uni-elem}.
In order to deduce the corresponding formulae we need to
substitute the expansions for plane waves
Eqs.~\eqref{eq:plane_iso} into Eqs.~\eqref{eq:wave_fun}
and make use of the matrix elements for the plane wave polarised along
the $z$ axis given by Eqs.~\eqref{eq:B.z-pl-m}-~\eqref{eq:B.z-pl-o}.
Note that Eqs.~\eqref{eq:iso} combined with
the orthogonality condition 
for the vector spherical harmonics~\eqref{eq:A.10}
immediately give the following relations:
\begin{widetext}
\begin{gather}
\langle\,
\vc{Y}_{jm}^{(m)\,*}(\uvc{r})\cdot
\vc{M}_{j'm'}^{(\beta)}(\rho,\uvc{r})
\,\rangle_{\uvc{r}}=\delta_{m,\alpha}\,
\delta_{jj'}\,\delta_{mm'}\, j_j(\rho)
\label{eq:aux.D.1}\\
\langle\,
\vc{Y}_{jm}^{(e)\,*}(\uvc{r})\cdot
\vc{M}_{j'm'}^{(\beta)}(\rho,\uvc{r})
\,\rangle_{\uvc{r}}=\delta_{e,\alpha}\,
\delta_{jj'}\,\delta_{mm'}\, Dj_j(\rho)\, .
\label{eq:aux.D.2}
\end{gather}
\end{widetext}

Below we write the resulting expressions for the matrix elements 
that correspond to transverse components of the wave
functions~\eqref{eq:wave_fun}.
\begin{widetext}
\begin{align}
&
p_{jj';\,m}^{\,(m,\,m)}(\rho)=
N_{jj'}\,
\langle\,
d^{\,(y;\,y)}_{j'j;\,m}\, j_j(\rho_e)
+ d^{\,(x;\,x)}_{j'j;\,m}\, j_j(\rho)
\,\rangle_{\theta_k}+
\Delta\,p_{jj';\,m}^{\,(m,\,m)}(\rho)
\, ,
\label{eq:D.1}\\
&
p_{jj';\,m}^{\,(m,\,e)}(\rho)= i\,
N_{jj'}\,
\langle\,
d^{\,(x;\,y)}_{j'j;\,m}\, j_j(\rho_e)
+ d^{\,(y;\,x)}_{j'j;\,m}\, j_j(\rho)
\,\rangle_{\theta_k}+
\Delta\,p_{jj';\,m}^{\,(m,\,e)}(\rho)
\, ,
\label{eq:D.2}\\
&
p_{jj';\,m}^{\,(e,\,m)}(\rho)= -i\,
N_{jj'}\,
\langle\,
d^{\,(y;\,x)}_{j'j;\,m}\, Dj_j(\rho_e)
+ d^{\,(x;\,y)}_{j'j;\,m}\, Dj_j(\rho)
\,\rangle_{\theta_k}+
\Delta\,p_{jj';\,m}^{\,(m,\,e)}(\rho)
\, ,
\label{eq:D.3}\\
&
p_{jj';\,m}^{\,(e,\,e)}(\rho)=
N_{jj'}\,
\langle\,
d^{\,(x;\,x)}_{j'j;\,m}\, Dj_j(\rho_e)
+ d^{\,(y;\,y)}_{j'j;\,m}\, Dj_j(\rho)
\,\rangle_{\theta_k}+
\Delta\,p_{jj';\,m}^{\,(e,\,e)}(\rho)
\, ,
\label{eq:D.4}\\
&
q_{jj';\,m}^{\,(m,\,m)}(\rho)=
N_{jj'}\,
\langle\,
d^{\,(x;\,x)}_{j'j;\,m}\, j_j(\rho_e)\,n_e^{-1}
+ d^{\,(y;\,y)}_{j'j;\,m}\, j_j(\rho)
\,\rangle_{\theta_k}\, ,
\label{eq:D.5}\\
&
q_{jj';\,m}^{\,(m,\,e)}(\rho)= i\,
N_{jj'}\,
\langle\,
d^{\,(y;\,x)}_{j'j;\,m}\, j_j(\rho_e)\,n_e^{-1}
+ d^{\,(x;\,y)}_{j'j;\,m}\, j_j(\rho)
\,\rangle_{\theta_k}\, ,
\label{eq:D.6}\\
&
q_{jj';\,m}^{\,(e,\,m)}(\rho)= -i\,
N_{jj'}\,
\langle\,
d^{\,(x;\,y)}_{j'j;\,m}\, Dj_j(\rho_e)\,n_e^{-1}
+ d^{\,(y;\,x)}_{j'j;\,m}\, Dj_j(\rho)
\,\rangle_{\theta_k}\, ,
\label{eq:D.7}\\
&
q_{jj';\,m}^{\,(e,\,e)}(\rho)= i\,
N_{jj'}\,
\langle\,
d^{\,(y;\,y)}_{j'j;\,m}\, Dj_j(\rho_e)\,n_e^{-1}
+ d^{\,(x;\,x)}_{j'j;\,m}\, Dj_j(\rho)
\,\rangle_{\theta_k}\, ,
\label{eq:D.8}
\end{align}

where $N_{jj'}\equiv\dfrac{i^{j'-j}}{8} [(2j+1)(2j'+1)]^{1/2}$
and
$d^{\,(\alpha;\,\beta)}_{jj';\,m}\equiv
d^{\,(\alpha)}_{jm}(\theta_k) d^{\,(\beta)}_{j'm}(\theta_k),\quad
\alpha,\,\beta\in\{x,\,y,\,z\}$.
The terms $\Delta\,p_{jj';\,m}^{\,(\alpha,\,\beta)}(\rho)$
are given by
\begin{align}
&
\Delta\,p_{jj';\,m}^{\,(m,\,m)}(\rho)=
N_{jj'}\,
\frac{2m}{\sqrt{j(j+1)}}\, r_{jj';\,m}^{\,(z,\,x)}(\rho)\, ,
\label{eq:D.9}\\
&
\Delta\,p_{jj';\,m}^{\,(m,\,e)}(\rho)=
N_{jj'}\,
\frac{2m}{\sqrt{j(j+1)}}\, r_{jj';\,m}^{\,(z,\,y)}(\rho)\, ,
\label{eq:D.10}\\
&
\Delta\,p_{jj';\,m}^{\,(e,\,m)}(\rho)=
N_{jj'}\,
\frac{-2\,i}{2j+1}\, 
\biggl\{
[j((j+1)^2-m^2)/(j+1)]^{1/2}\,
r_{j+1\,j';\,m}^{\,(z,\,y)}(\rho)-\notag\\
&
\phantom{
\Delta\,p_{jj';\,m}^{\,(e,\,m)}(\rho)=
N_{jj'}\,
\frac{-2\,i}{2j+1}\,
}
-[(j+1)(j^2-m^2)/j]^{1/2}\,
r_{j-1\,j';\,m}^{\,(z,\,y)}(\rho)
\biggr\}\, ,
\label{eq:D.11}\\
&
\Delta\,p_{jj';\,m}^{\,(e,\,e)}(\rho)=
N_{jj'}\,
\frac{2}{2j+1}\, 
\biggl\{
[j((j+1)^2-m^2)/(j+1)]^{1/2}\,
r_{j+1\,j';\,m}^{\,(z,\,x)}(\rho)-\notag\\
&
\phantom{
\Delta\,p_{jj';\,m}^{\,(e,\,m)}(\rho)=
N_{jj'}\,
\frac{-2\,i}{2j+1}\,
}
-[(j+1)(j^2-m^2)/j]^{1/2}\,
r_{j-1\,j';\,m}^{\,(z,\,x)}(\rho)
\biggr\}\, ,
\label{eq:D.12}
\end{align}
where
\begin{equation}
r_{j\,j';\,m}^{\,(z,\,\alpha)}(\rho)=\frac{u}{u+1}
\langle\,
d^{\,(\alpha;\,z)}_{j'j;\,m}(\theta_k)
\, j_j(\rho_e)\,
\sin\theta_k
\,\rangle_{\theta_k}\, .
\label{eq:D.13}
\end{equation}
\end{widetext}
Expressions for $\tilde{p}_{jj';\,m}^{\,(\alpha,\,\beta)}(\rho)$ 
and $\tilde{q}_{jj';\,m}^{\,(\alpha,\,\beta)}(\rho)$ 
can be derived from~Eqs.~\eqref{eq:D.1}--\eqref{eq:D.13} by replacing
spherical Bessel functions of the first kind $j_j(x)$
with spherical Bessel functions of the third kind
$h_j^{(1)}(x)$.

The coefficient functions 
$p_{jj';\,m}^{\,(\alpha,\,\beta)}$ and $q_{jj';\,m}^{\,(\alpha,\,\beta)}$
are diagonal, $\propto\delta_{\alpha,\,\beta}\,\delta_{jj'}$, 
in the limit of weak anisotropy, $u\to 0$.
It can be seen from orthogonality
condition~\eqref{eq:A.9} that
provides the orthogonality conditions for $d$-functions
defined by Eqs.~\eqref{eq:d_x}-\eqref{eq:d_y_z}
in the following form
\begin{gather}
\langle\,
d^{\,(x;\,x)}_{j'j;\,m}
+ d^{\,(y;\,y)}_{j'j;\,m}
\,\rangle_{\theta_k}=\frac{8}{2j+1}\,\delta_{jj'}\, ,\notag\\
\langle\,
d^{\,(x;\,y)}_{j'j;\,m}
+ d^{\,(y;\,x)}_{j'j;\,m}
\,\rangle_{\theta_k}=0\, .
\label{eq:D.14}
\end{gather}


\begin{thebibliography}{29}
\expandafter\ifx\csname natexlab\endcsname\relax\def\natexlab#1{#1}\fi
\expandafter\ifx\csname bibnamefont\endcsname\relax
  \def\bibnamefont#1{#1}\fi
\expandafter\ifx\csname bibfnamefont\endcsname\relax
  \def\bibfnamefont#1{#1}\fi
\expandafter\ifx\csname citenamefont\endcsname\relax
  \def\citenamefont#1{#1}\fi
\expandafter\ifx\csname url\endcsname\relax
  \def\url#1{\texttt{#1}}\fi
\expandafter\ifx\csname urlprefix\endcsname\relax\def\urlprefix{URL }\fi
\providecommand{\bibinfo}[2]{#2}
\providecommand{\eprint}[2][]{\url{#2}}

\bibitem[{\citenamefont{Mie}(1908)}]{Mie:1908}
\bibinfo{author}{\bibfnamefont{G.}~\bibnamefont{Mie}}, \bibinfo{journal}{Ann.
  Phys. (Leipzig)} \textbf{\bibinfo{volume}{25}}, \bibinfo{pages}{377}
  (\bibinfo{year}{1908}).

\bibitem[{\citenamefont{Asano and Yamamoto}(1975)}]{Asan:1975}
\bibinfo{author}{\bibfnamefont{S.}~\bibnamefont{Asano}} \bibnamefont{and}
  \bibinfo{author}{\bibfnamefont{G.}~\bibnamefont{Yamamoto}},
  \bibinfo{journal}{Appl. Opt.} \textbf{\bibinfo{volume}{14}},
  \bibinfo{pages}{29} (\bibinfo{year}{1975}).

\bibitem[{\citenamefont{Roth and Digman}(1973)}]{Rot:1973}
\bibinfo{author}{\bibfnamefont{J.}~\bibnamefont{Roth}} \bibnamefont{and}
  \bibinfo{author}{\bibfnamefont{M.}~\bibnamefont{Digman}},
  \bibinfo{journal}{J. Opt. Soc. Am.} \textbf{\bibinfo{volume}{63}},
  \bibinfo{pages}{308} (\bibinfo{year}{1973}).

\bibitem[{\citenamefont{Lange and Aragon}(1990)}]{Arag:1990}
\bibinfo{author}{\bibfnamefont{B.}~\bibnamefont{Lange}} \bibnamefont{and}
  \bibinfo{author}{\bibfnamefont{S.}~\bibnamefont{Aragon}},
  \bibinfo{journal}{J. Chem. Phys.} \textbf{\bibinfo{volume}{92}},
  \bibinfo{pages}{4643} (\bibinfo{year}{1990}).

\bibitem[{\citenamefont{Hahn and Aragon}(1994)}]{Arag:1994}
\bibinfo{author}{\bibfnamefont{D.}~\bibnamefont{Hahn}} \bibnamefont{and}
  \bibinfo{author}{\bibfnamefont{S.}~\bibnamefont{Aragon}},
  \bibinfo{journal}{J. Chem. Phys.} \textbf{\bibinfo{volume}{101}},
  \bibinfo{pages}{8409} (\bibinfo{year}{1994}).

\bibitem[{\citenamefont{Karacali et~al.}(1997)\citenamefont{Karacali, Risseler,
  and Ferris}}]{Kar:1997}
\bibinfo{author}{\bibfnamefont{H.}~\bibnamefont{Karacali}},
  \bibinfo{author}{\bibfnamefont{S.}~\bibnamefont{Risseler}}, \bibnamefont{and}
  \bibinfo{author}{\bibfnamefont{K.}~\bibnamefont{Ferris}},
  \bibinfo{journal}{Phys. Rev. B} \textbf{\bibinfo{volume}{56}},
  \bibinfo{pages}{4286} (\bibinfo{year}{1997}).

\bibitem[{\citenamefont{Kiselev
  et~al.}(2000{\natexlab{a}})\citenamefont{Kiselev, Reshetnyak, and
  Sluckin}}]{Kis:2000:biano}
\bibinfo{author}{\bibfnamefont{A.}~\bibnamefont{Kiselev}},
  \bibinfo{author}{\bibfnamefont{V.}~\bibnamefont{Reshetnyak}},
  \bibnamefont{and} \bibinfo{author}{\bibfnamefont{T.}~\bibnamefont{Sluckin}},
  in \emph{\bibinfo{booktitle}{Proc. Bianisotropics'2000}}, edited by
  \bibinfo{editor}{\bibfnamefont{A.}~\bibnamefont{Barbosa}} \bibnamefont{and}
  \bibinfo{editor}{\bibfnamefont{A.}~\bibnamefont{Topa}}
  (\bibinfo{publisher}{8th Intern. Conf. on Electromagnetics of Complex Media},
  \bibinfo{address}{Lisbon, Portugal}, \bibinfo{year}{2000}{\natexlab{a}}), pp.
  \bibinfo{pages}{343--346}.

\bibitem[{\citenamefont{Kiselev
  et~al.}(2000{\natexlab{b}})\citenamefont{Kiselev, Reshetnyak, and
  Sluckin}}]{Kis:2000:opt}
\bibinfo{author}{\bibfnamefont{A.}~\bibnamefont{Kiselev}},
  \bibinfo{author}{\bibfnamefont{V.}~\bibnamefont{Reshetnyak}},
  \bibnamefont{and} \bibinfo{author}{\bibfnamefont{T.}~\bibnamefont{Sluckin}},
  \bibinfo{journal}{Opt. Spectrosc.}
  \textbf{\bibinfo{volume}{89}}(\bibinfo{number}{6}), \bibinfo{pages}{907}
  (\bibinfo{year}{2000}{\natexlab{b}}).

\bibitem[{\citenamefont{Newton}(1982)}]{New}
\bibinfo{author}{\bibfnamefont{R.}~\bibnamefont{Newton}},
  \emph{\bibinfo{title}{Scattering Theory of Waves and Particles}}
  (\bibinfo{publisher}{Springer}, \bibinfo{address}{Heidelberg},
  \bibinfo{year}{1982}), \bibinfo{edition}{2nd} ed.

\bibitem[{\citenamefont{Ishimaru}(1978)}]{Ishim}
\bibinfo{author}{\bibfnamefont{A.}~\bibnamefont{Ishimaru}},
  \emph{\bibinfo{title}{Wave Propagation and Scattering in Random Media}}
  (\bibinfo{publisher}{Academic Press}, \bibinfo{address}{New York},
  \bibinfo{year}{1978}).

\bibitem[{\citenamefont{Kreuzer and Eidenschink}(1996)}]{Kre:1996}
\bibinfo{author}{\bibfnamefont{M.}~\bibnamefont{Kreuzer}} \bibnamefont{and}
  \bibinfo{author}{\bibfnamefont{R.}~\bibnamefont{Eidenschink}}, in
  \emph{\bibinfo{booktitle}{Liquid Crystals in Complex Geometries}}, edited by
  \bibinfo{editor}{\bibfnamefont{G.}~\bibnamefont{Crawford}} \bibnamefont{and}
  \bibinfo{editor}{\bibfnamefont{S.}~\bibnamefont{\v{Z}umer}}
  (\bibinfo{publisher}{Taylor \& Francis}, \bibinfo{address}{London},
  \bibinfo{year}{1996}), chap.~\bibinfo{chapter}{15}.

\bibitem[{\citenamefont{Bellini et~al.}(1998)\citenamefont{Bellini, Clark,
  Degiorgio, Mantegazza, and Natale}}]{Bel:1998}
\bibinfo{author}{\bibfnamefont{T.}~\bibnamefont{Bellini}},
  \bibinfo{author}{\bibfnamefont{N.}~\bibnamefont{Clark}},
  \bibinfo{author}{\bibfnamefont{V.}~\bibnamefont{Degiorgio}},
  \bibinfo{author}{\bibfnamefont{F.}~\bibnamefont{Mantegazza}},
  \bibnamefont{and} \bibinfo{author}{\bibfnamefont{G.}~\bibnamefont{Natale}},
  \bibinfo{journal}{Phys. Rev. E} \textbf{\bibinfo{volume}{57}},
  \bibinfo{pages}{2996} (\bibinfo{year}{1998}).

\bibitem[{\citenamefont{Mishchenko et~al.}(1996)\citenamefont{Mishchenko,
  Travis, and Mackowski}}]{Mis:1996}
\bibinfo{author}{\bibfnamefont{M.}~\bibnamefont{Mishchenko}},
  \bibinfo{author}{\bibfnamefont{L.}~\bibnamefont{Travis}}, \bibnamefont{and}
  \bibinfo{author}{\bibfnamefont{D.}~\bibnamefont{Mackowski}},
  \bibinfo{journal}{J. of Quant. Spectr. {\upshape\&} Radiat. Transf.}
  \textbf{\bibinfo{volume}{55}}, \bibinfo{pages}{535} (\bibinfo{year}{1996}).

\bibitem[{\citenamefont{Mishchenko et~al.}(2000)\citenamefont{Mishchenko,
  Hovenier, and Travis}}]{Mish}
\bibinfo{editor}{\bibfnamefont{M.}~\bibnamefont{Mishchenko}},
  \bibinfo{editor}{\bibfnamefont{J.}~\bibnamefont{Hovenier}}, \bibnamefont{and}
  \bibinfo{editor}{\bibfnamefont{L.}~\bibnamefont{Travis}}, eds.,
  \emph{\bibinfo{title}{Light Scattering by Nonspherical Particles: Theory,
  Measurements and Applications}} (\bibinfo{publisher}{Academic Press},
  \bibinfo{address}{New York}, \bibinfo{year}{2000}).

\bibitem[{\citenamefont{Boren and Huffman}(1983)}]{Bor}
\bibinfo{author}{\bibfnamefont{C.}~\bibnamefont{Boren}} \bibnamefont{and}
  \bibinfo{author}{\bibfnamefont{D.}~\bibnamefont{Huffman}},
  \emph{\bibinfo{title}{Absorption and Scattering of Light by Small Particles}}
  (\bibinfo{publisher}{Wiley-Interscience}, \bibinfo{address}{New York},
  \bibinfo{year}{1983}).

\bibitem[{\citenamefont{\v{Z}umer and Doane}(1986)}]{Zum:1986}
\bibinfo{author}{\bibfnamefont{S.}~\bibnamefont{\v{Z}umer}} \bibnamefont{and}
  \bibinfo{author}{\bibfnamefont{J.}~\bibnamefont{Doane}},
  \bibinfo{journal}{Phys. Rev. A} \textbf{\bibinfo{volume}{34}},
  \bibinfo{pages}{3373} (\bibinfo{year}{1986}).

\bibitem[{\citenamefont{\v{Z}umer}(1988)}]{Zum:1988}
\bibinfo{author}{\bibfnamefont{S.}~\bibnamefont{\v{Z}umer}},
  \bibinfo{journal}{Phys. Rev. A} \textbf{\bibinfo{volume}{37}},
  \bibinfo{pages}{4006} (\bibinfo{year}{1988}).

\bibitem[{\citenamefont{Biedenharn and Louck}(1981)}]{Bie}
\bibinfo{author}{\bibfnamefont{L.}~\bibnamefont{Biedenharn}} \bibnamefont{and}
  \bibinfo{author}{\bibfnamefont{J.}~\bibnamefont{Louck}},
  \emph{\bibinfo{title}{Angular Momentum in Quantum Physics}}
  (\bibinfo{publisher}{Addison--Wesley}, \bibinfo{address}{Reading,
  Massachusetts}, \bibinfo{year}{1981}).

\bibitem[{\citenamefont{Abramowitz and Stegun}(1972)}]{Abr}
\bibinfo{editor}{\bibfnamefont{M.}~\bibnamefont{Abramowitz}} \bibnamefont{and}
  \bibinfo{editor}{\bibfnamefont{I.}~\bibnamefont{Stegun}}, eds.,
  \emph{\bibinfo{title}{Handbook of Mathematical Functions}}
  (\bibinfo{publisher}{Dover}, \bibinfo{address}{New York},
  \bibinfo{year}{1972}).

\bibitem[{\citenamefont{Gelfand et~al.}(1963)\citenamefont{Gelfand, Minlos, and
  Shapiro}}]{Gel}
\bibinfo{author}{\bibfnamefont{I.}~\bibnamefont{Gelfand}},
  \bibinfo{author}{\bibfnamefont{R.}~\bibnamefont{Minlos}}, \bibnamefont{and}
  \bibinfo{author}{\bibfnamefont{Z.}~\bibnamefont{Shapiro}},
  \emph{\bibinfo{title}{Representations of Rotation and Lorenz Groups and Their
  Applications}} (\bibinfo{publisher}{Pergamon Press},
  \bibinfo{address}{Oxford}, \bibinfo{year}{1963}).

\bibitem[{\citenamefont{Born and Wolf}(1980)}]{Born}
\bibinfo{author}{\bibfnamefont{M.}~\bibnamefont{Born}} \bibnamefont{and}
  \bibinfo{author}{\bibfnamefont{E.}~\bibnamefont{Wolf}},
  \emph{\bibinfo{title}{Principles of Optics}} (\bibinfo{publisher}{Pergamon
  Press}, \bibinfo{address}{Oxford}, \bibinfo{year}{1980}),
  \bibinfo{edition}{2nd} ed.

\bibitem[{\citenamefont{Landau and Lifshitz}(1984)}]{Landau:el}
\bibinfo{author}{\bibfnamefont{L.}~\bibnamefont{Landau}} \bibnamefont{and}
  \bibinfo{author}{\bibfnamefont{E.}~\bibnamefont{Lifshitz}},
  \emph{\bibinfo{title}{Elecrtodynamics of Continuous Media}}
  (\bibinfo{publisher}{Pergamon}, \bibinfo{address}{Oxford},
  \bibinfo{year}{1984}).

\bibitem[{\citenamefont{Lax and Nelson}(1971)}]{Lax:1971}
\bibinfo{author}{\bibfnamefont{M.}~\bibnamefont{Lax}} \bibnamefont{and}
  \bibinfo{author}{\bibfnamefont{D.}~\bibnamefont{Nelson}},
  \bibinfo{journal}{Phys. Rev. B} \textbf{\bibinfo{volume}{4}},
  \bibinfo{pages}{3694} (\bibinfo{year}{1971}).

\bibitem[{\citenamefont{Stark and Lubensky}(1997)}]{Stark:1997}
\bibinfo{author}{\bibfnamefont{H.}~\bibnamefont{Stark}} \bibnamefont{and}
  \bibinfo{author}{\bibfnamefont{T.}~\bibnamefont{Lubensky}},
  \bibinfo{journal}{Phys. Rev. E} \textbf{\bibinfo{volume}{55}},
  \bibinfo{pages}{514} (\bibinfo{year}{1997}).

\bibitem[{\citenamefont{Lacoste et~al.}(1998)\citenamefont{Lacoste, van
  Tiggelen, Rikken, and Sparenberg}}]{Lacos:1998}
\bibinfo{author}{\bibfnamefont{D.}~\bibnamefont{Lacoste}},
  \bibinfo{author}{\bibfnamefont{B.}~\bibnamefont{van Tiggelen}},
  \bibinfo{author}{\bibfnamefont{G.}~\bibnamefont{Rikken}}, \bibnamefont{and}
  \bibinfo{author}{\bibfnamefont{A.}~\bibnamefont{Sparenberg}},
  \bibinfo{journal}{J. Opt. Soc. Am. A} \textbf{\bibinfo{volume}{15}},
  \bibinfo{pages}{1636} (\bibinfo{year}{1998}).

\bibitem[{\citenamefont{van Tiggelen}(1995)}]{Tigg:1995}
\bibinfo{author}{\bibfnamefont{B.}~\bibnamefont{van Tiggelen}},
  \bibinfo{journal}{Phys. Rev. Lett.} \textbf{\bibinfo{volume}{75}},
  \bibinfo{pages}{422} (\bibinfo{year}{1995}).

\bibitem[{\citenamefont{Rikken and van Tiggelen}(1996)}]{Tigg1:1996}
\bibinfo{author}{\bibfnamefont{G.}~\bibnamefont{Rikken}} \bibnamefont{and}
  \bibinfo{author}{\bibfnamefont{B.}~\bibnamefont{van Tiggelen}},
  \bibinfo{journal}{Nature} \textbf{\bibinfo{volume}{381}}, \bibinfo{pages}{54}
  (\bibinfo{year}{1996}).

\bibitem[{\citenamefont{Jing et~al.}(1992)\citenamefont{Jing, Sheng, and
  Zhou}}]{Sheng:1992}
\bibinfo{author}{\bibfnamefont{X.}~\bibnamefont{Jing}},
  \bibinfo{author}{\bibfnamefont{P.}~\bibnamefont{Sheng}}, \bibnamefont{and}
  \bibinfo{author}{\bibfnamefont{M.}~\bibnamefont{Zhou}},
  \bibinfo{journal}{Phys. Rev. A} \textbf{\bibinfo{volume}{46}},
  \bibinfo{pages}{6513} (\bibinfo{year}{1992}).

\bibitem[{\citenamefont{Soukoulis et~al.}(1994)\citenamefont{Soukoulis, Datta,
  and Economou}}]{Souk:1994a}
\bibinfo{author}{\bibfnamefont{C.}~\bibnamefont{Soukoulis}},
  \bibinfo{author}{\bibfnamefont{S.}~\bibnamefont{Datta}}, \bibnamefont{and}
  \bibinfo{author}{\bibfnamefont{E.}~\bibnamefont{Economou}},
  \bibinfo{journal}{Phys. Rev. B} \textbf{\bibinfo{volume}{49}},
  \bibinfo{pages}{3800} (\bibinfo{year}{1994}).

\end{thebibliography}

\end{document}